\documentclass[12pt,preprint]{aastex}

\newcommand{\beq}{\begin{equation}}
\newcommand{\eeq}{\end{equation}}

\shorttitle{OVI in Elliptical Galaxies}
\shortauthors{Bregman et al.}

\begin{document}

\title{Mass Loss From Evolved Stars in Elliptical Galaxies}

\author{Joel R. Parriott and Joel N. Bregman}

\affil{Department of Astronomy, University of Michigan, Ann Arbor, MI 48109}
\email{jbregman@umich.edu}

\begin{abstract}
Most of the X-ray emitting gas in early-type galaxies probably 
originates from red giant mass loss and here we model the interaction 
between this stellar mass loss and the hot ambient medium.  Using 
two-dimensional hydrodynamic simulations, we adopt a temperature 
for the ambient medium of $3 \times 10^6 $ K along with a range of 
ambient densities and stellar velocities.  When the stellar velocity 
is supersonic relative to the ambient medium, a bow shock occurs, 
along with a shock driven into the stellar ejecta, which heats only 
a fraction of the gas.  Behind the bow shock, a cool wake develops 
but the fast flow of the hot medium causes Kelvin-Helmholtz 
instabilities to grow and these fingers are shocked and heated (without 
radiative cooling).  Along with the mixing of this wake material 
with the hot medium, most of the stellar ejecta is heated to 
approximately the temperature of the hot ambient medium within 2 pc 
of the star.  With the addition of radiative cooling, some wake 
material remains cool ($< 10^5$ K), accounting for up to 25\% of 
the stellar mass loss.  Less cooled gas survives when the ambient 
density is lower or when the stellar velocity is higher than in our 
reference case. These results suggest that some cooled gas should 
be present in the inner part of early-type galaxies that have a hot 
ambient medium.  These calculations may explain the observed 
distributed optical emission line gas as well as the presence of 
dust in early-type galaxies.

\end{abstract}

\keywords{galaxies: ISM ---- cooling flows ---- X-rays: galaxies ---- stars: mass loss}

\section{Introduction}

The observed metallicities of hot gas in early-type galaxies disagree with
elementary predictions.  The source of the hot galactic gas is expected to
be the mass lost by evolved stars.  The main contribution to the gas mass
is from giant stars and planetary nebulae and this gas initially has the
velocity of the parent star.  As this material collides with the mass lost
from other stars or with the hot ambient medium within the galaxy, the
mass loss material is shocked to a temperature corresponding to the
velocity dispersion of the stars in the galaxy (review by \citealt{math03}).  
If this were the only heating
mechanism, the gas would have the velocity dispersion temperature and a
metallicity of the stars.  The metallicity of early-type galaxies have been
measured, and for galaxies near L$_{*}$, the metallicity is approximately solar
within R$_e$/2 \citep{trag00a}.

However, there is a second source of metals, due to the occurrence of
Type Ia supernovae in these galaxies \citep{tcb94}.  The mass of Fe from
these supernovae is many times greater than that in the mass lost from
giant stars, so the net result is that the Fe abundance of the hot ISM in
early-type galaxies should be 3-10 times the solar value \citep{renz93,brig05}.  The other
prediction is that the heating by the supernovae should raise the gas
temperature above that due only to the velocity dispersion of the stars. 
While the gas temperature is indeed elevated above the velocity
dispersion temperature (e.g., \citealt{bb98}), the metallicities are far from these predicted
values, lying near the Solar value in the more X-ray luminous early-type
galaxies \citep{athey03,hump06}.

One modification to the model is that gas from the surrounding
environment has fallen into the galaxy, diluting the metallicity of the gas \citep{math03}. 
This might predict a sharp radial metallicity gradient, with more dilution
in the outer parts and less in the inner part of the galaxy, where there is a
lower ratio of the stellar mass loss rate to the density of the group or
cluster medium.  Such a gradient is not generally detected in the X-ray
observations of the hot gas \citep{athey03,hump06}.

The issue of external gas accretion should be relatively unimportant in
galaxies that are thought to be driving galactic winds.  Partial or total
galactic winds are the leading explanation for the X-ray poor early-type
galaxies, whose gas content is an order of magnitude less than their X-ray
bright counterparts.  In these systems, the supernova heating may be
enhanced, thereby driving the galactic wind.  In this picture, \citet{david06}
estimate that the Fe abundance would be an order of magnitude
above the Solar value.  However, observations of X-ray faint early-type
galaxies shows that the Fe abundance in the hot gas is lower than the 
X-ray bright ellipticals, with typical values that are 0.3 of the Solar value \citep{irwin06}. 
Evidently, there is a metallicity problem: the metals in the mass lost by
stars is not effectively mixed into the host ISM of these galaxies.

This metallicity problem has been recognized for over a decade (e.g.,
\citealt{fuji96,fuji97}) and a resolution to the problem may be that
the metals do not fully mix into the ISM \citep{math90,brig05}.  One aspect of 
this is that not all stars have the same metallicity.  As these authors 
point out, higher metallicity gas radiates more effectively, so it might
cool to T $\leq$ 10$^{{\rm 4}}$ K if the cooling time is shorter than the heating and
mixing timescale.  This would produce a bias against high-metallicity gas
becoming part of the hot ISM.  Estimations for the magnitude of the
effect have been calculated, which involve spherically symmetric bubbles and
shocks that propagate into the mass lost from stars.  However,
the more realistic situation will have some important differences.  For the
typical condition where the star is losing mass as it moves through the
ambient ISM at a few hundred km s$^{-1}$, there will be a bow shock in front of
the star.  As the shocked gas flows by the stellar mass loss, it will create a
wake with strong velocity gradients.  Kelvin-Helmholtz instabilities are
likely to develop between the rapidly flowing shocked gas and the stellar
mass loss, and this will lead to shocking of the cooler gas and mixing
between the two components.  These processes are difficult to estimate
accurately, so we have undertaken a series of numerical fluid dynamical
calculations to study the process by which stellar mass loss interacts with
the hot ambient medium.

\section{Computational Methodology}

The gas dynamics code used in this work is based upon the highly accurate
Piecewise Parabolic Method (PPM) of \citet{cole84}.  
The choice of PPM as the numerical scheme for these simulations is based upon 
the hydrodynamics that are expected to be important in our physical problem.
PPM is robust, and its high spatial and temporal accuracy (discontinuity 
capturing) capabilities make it a natural choice, since vortex mixing and 
shock heating are expected to be the key mechanisms for transforming the 
cold stellar ejecta into hot ambient gas.
The implementation of PPM used here \citep{blond93} contains improvements 
that make it more suitable for the geometry and boundary conditions found 
in our problem.  The PPM scheme of \citet{cole84}
was written to operate on a grid defined by the volume $V(r)$ rather than
the position $r$ (remap step) or mass coordinate $m$ (Lagrangian step),
thus rendering the interpolation Cartesian in $V$. 
In this way, curvilinear geometry can be incorporated without having to
correct for coordinate singularities.  However, a volume-based interpolation
does not accurately capture the solution on a reflecting symmetry boundary,
where the velocity $u$ is generally linear in $r$, not $V(r)$.  So even
though the original PPM scheme can account for curvilinear geometries, the
cumulative truncation error for advection of a velocity interpolated on
the volume coordinate near a symmetry axis is first-order in $\Delta r/r$.
\citet{blond93} demonstrated that the PPM-LR scheme retains greater
accuracy near coordinate singularities when using interpolations in $r$
(or $\theta$) instead of $V(r)$.
Three relatively simple modifications to the original scheme are required
for improved accuracy in both the radial and azimuthal coordinates.

We began with the serial version of the code of \citet{blond93}, known as 
VH-1 and written for distribution by the Virginia Institute of Theoretical 
Astrophysics (VITA) Numerical Astrophysics Group.  This is a two-dimensional
code that we use in spherical polar coordinates.  The VH-1 code
implements PPM as a Lagrangian hydrodynamic step coupled with a remap
onto the original fixed Eulerian grid (referred to as PPM-LR).  This scheme
combines the advantages of Lagrangian and Eulerian methods by solving the
fluid equations in Lagrangian mode and then remapping the dynamical variables
back to the fixed Eulerian grid. Because the physics and the
coordinate transformation are solved for separately.  If the remap and the 
Lagrangian scheme are each third order, the entire solution will be third order.

The original VH-1 code was modified in a few significant ways before it could
be used to address our problem.  First, we introduced optically thin radiative
losses, which was implemented by the operator splitting method outside of the Riemann
solver after each full time step; for further discussion of the VH-1 code and
the modifications, see \citet{parr98}.  While this represents a loss of
self-consistency with the gas dynamics, it is a much used and validated
method of including heating and/or cooling.  We use the local, time-independent
collisional ionization equilibrium cooling function, $\Lambda_{N}(T)$, of
\citep{sd93}.  These tables cover a temperature range of $10^{4} K - 10^{8.5} K$,
and a large metallicity range of $[Fe/H] = -3.0 - +0.5$.

The routine employs an implicit scheme, the secant method, to
calculate the energy (or equivalently, the pressure) lost in one time step.
The computational
load incurred by an implicit method is justified by its ability to minimize
the small errors introduced when removing energy from the system.  In practice,
the code uses interpolated values from a table of radiative cooling function
values as a function of temperature, and iterates toward a value for the
change in temperature for each cell.  The iteration is carried out a maximum
of twelve times, or until the fractional temperature change is less than
one part in a thousand, $\Delta T/T < 10^{-3}$.  A new temperature is
calculated for each cell, and the pressure is changed accordingly (in this
case via the ideal gas equation of state):

\beq
\label{eq:elost}
\Delta p = \frac{\rho\,k_b \Delta T}{\mu m_p}\:,\;{\rm where}\;\; \Delta T = \left( \frac{\gamma - 1}{4 \mu m_{p} k_b} \right) \Lambda_{N} \rho\, \Delta t \,.
\eeq

Finally, we tested the output of the cooling routine
using a flat cooling function and a single temperature grid with no inflow or
outflow.  The temperature drop over the time taken by $N$ time steps had to
exactly match analytical expectations.

\begin{figure}
\plotone{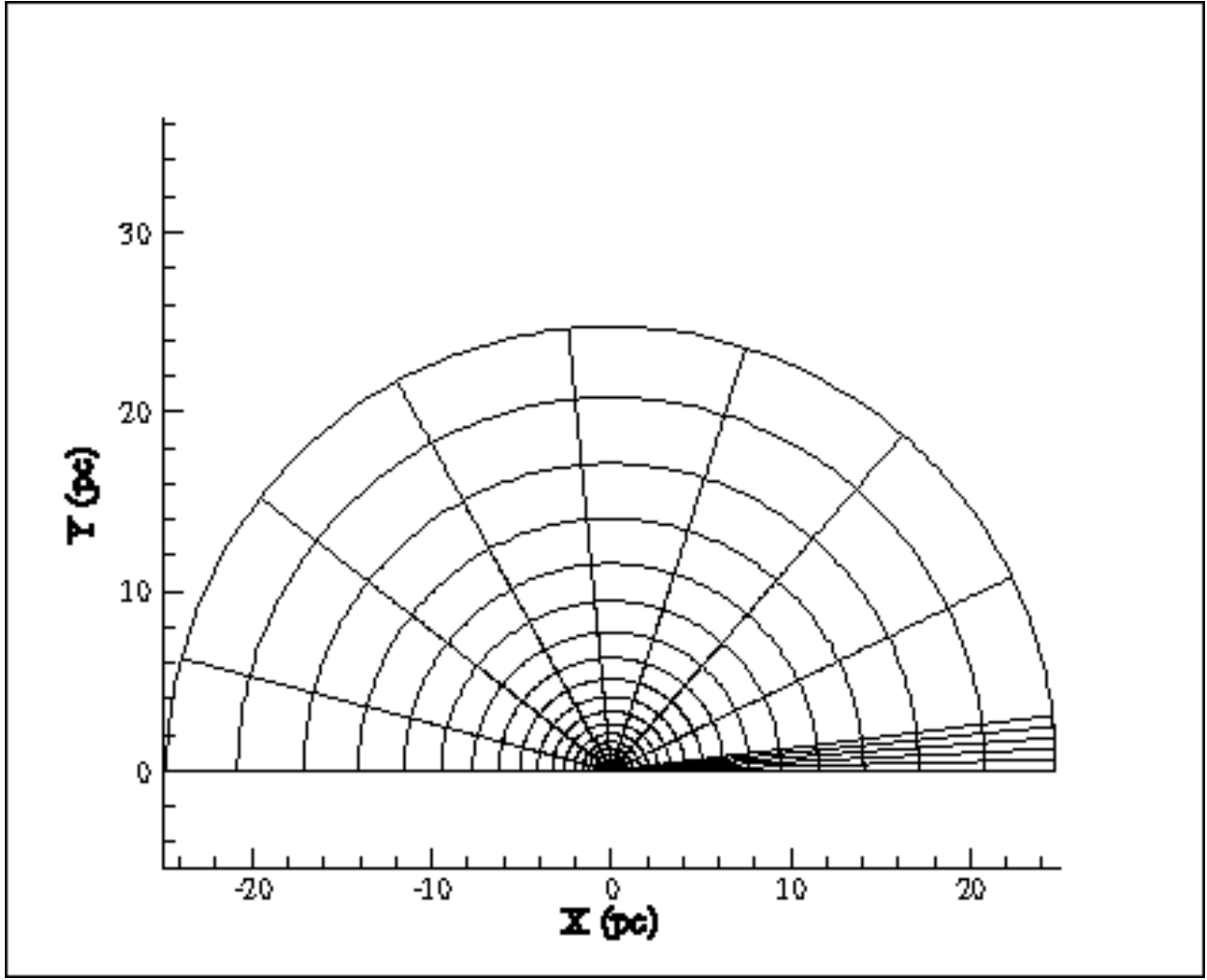}
\caption[The 2-D axisymmetric spherical polar grid mesh]%
        {The 2-D axisymmetric spherical polar grid mesh used for every 
	simulation in this work.  The resolution, determined by the spacing of 
	the radial and angular cells, is highest near the center and in the 
	downstream wake region.  For clarity, this figure only shows 
	the radial and angular cells in 20-cell increments (there are 
	360 radial cells and 240 angular cells).  The polar axis is horizontal
	in this figure; further details are given in the text.}
\label{fig:mesh}
\end{figure}

The second change was to the computational grid.
We maintained the regular nature of the 2D spherical polar grid,
but created an unequal spacing of grid cells.  
The outermost radius is 25 pc and at the innermost radius, 0.01 pc,
the stellar surface is not resolved.  The radial length of the
cells increase by a factor of 1.01 per cell, which creates much higher radial
resolution in the inner grid.  This growth dictates an outermost cell
$\approx 36$ times longer than the innermost cell when using 360 radial cells.  
This inner resolution is important since it is
the region where the instabilities begin to grow.  We also found it important
in helping the lower energy density material flow into the grid from the
inner boundary without encountering non-physical negative pressures in several
cells near the boundary.  Because most of the important structure develops
in the wake along the downstream symmetry polar axis, we also increased the azimuthal
resolution by a factor of 10 for the cells in that region (Figure~\ref{fig:mesh}).  This is
accomplished by calculating a beginning angular cell size, decreasing
that size by an order of magnitude and forming 100 azimuthal cells at that
resolution.  The remaining cells fill in the rest of the $\pi$ radians to
complete the grid.

The third modification concerns the boundary procedures for the 2D axisymmetric
spherical polar domain.  The outer boundary of the grid needed to have an
inflow boundary condition on the left side for the hot, fast ambient wind,
and an outflow only condition on the right side for the gas to leave the grid.
Since the public VH-1 code treats the entire outer boundary as one boundary
with one condition, we split the boundary in half.  A particular 1D radial
sweep uses the inflow condition if its angular theta value lies in the second
quadrant ($cos(\theta) < 0$), or uses the outflow condition if its theta is in
the first quadrant ($cos(\theta) \geq 0$).  The inflow boundary condition
was set by zero-gradient fixed values for the pressure, density, velocity,
and derived total energy of the hot, fast ambient wind.
We also used a zero-gradient
outflow condition, but added a pressure condition which set the minimum
pressure of the outflow ghost zones at the fixed inflow pressure value.  While
this change is unnecessary for supersonic flows, it helps to reduce the influence
of sound waves hitting the boundary for sub-sonic flows, and prevents any
pressure buildup near the outflow boundary.  A dissertation on boundary 
procedures for fluid dynamics found that
boundary conditions, no matter how well-conceived and physically elegant, can 
never eliminate all of the problems incurred at outer boundaries, and one should
make the boundary as far away as practically possible (\citealt{body92}).

The boundaries along the upwind and downwind symmetry axis were implemented 
with reflecting boundary conditions.  The density and pressure gradients are 
set to zero at the boundary, and the sign of the velocity is changed.

\begin{figure}
\plotone{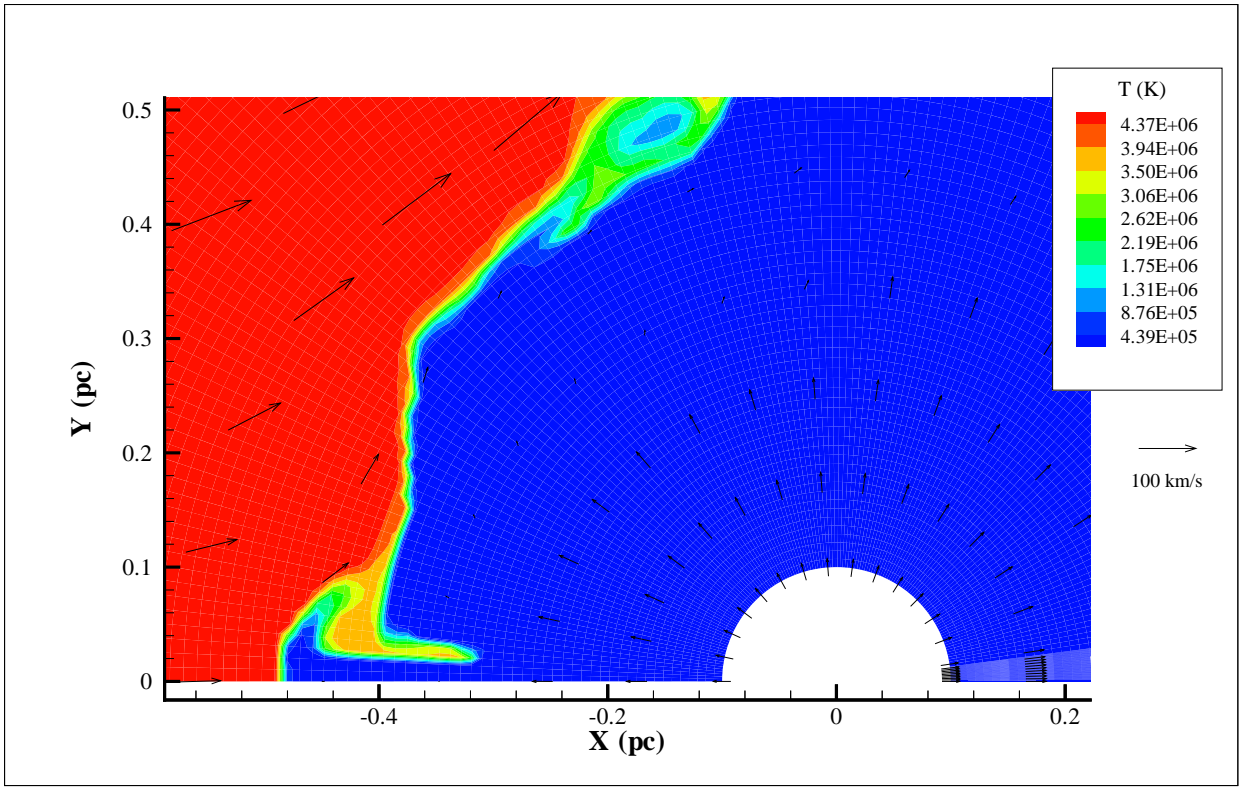}
\caption[Example of stellar wind upstream symmetry axis errors]%
        {Example of upstream extension of gas along symmetry axis, shown
	with the temperature map.  This small amount of material is
	found $0.04$ parsecs further upstream than the contact discontinuity 
	for the stellar wind.}
\label{fig:leakage}
\end{figure}

The inner boundary adjacent to the upstream symmetry axis was troublesome in
that a small amount of material representing the stellar
outflow could work its way upstream further than the bulk of the ram pressure
confined stellar wind (see Figure~\ref{fig:leakage}).
Although VH-1 includes the non-Cartesian improvements to the spatial
interpolation routine, it does not 
completely eliminate the errors \citep{blond93}.  The restriction of 
axisymmetry makes this axis a favorable place for such errors to grow,
even though the improved interpolation is implemented (\citealt{bpriv98}).  
This behavior in VH-1 was noted by \citet{dinge97}, where it occurred 
in his simulations of proto-globular cluster clouds.
After a number of attempts to fix this problem, we conclude that it
is difficult to remove without moving to a three-dimensional code.  
However, the error that it introduces does not appear to affect the 
behavior and evolution of the wake, which is orders of magnitude more
massive and energetically important.

The final change to the serial version of VH-1 involved a test and correction 
if the pressure of a given cell becomes negative during the remap step.  
This happens when the kinetic
energy $u^2$ and thermal energy $NkT$ of a cell are similar in magnitude, so 
that small errors in the velocity can drive the derived thermal pressure 
negative ($p_{derived} \propto \rho (E_{thermal} - E_{kinetic})$).  We repair
the problem cell by computing a new pressure from the average temperature of
the adjacent two cells in the direction of the current sweep.  We are injecting
energy into the grid using this method, but the amounts of energy are very small,
so the effects on the general flow are minor.

The public version of VH-1 was a serial code, which for our problem, 
would have taken too long to produce results for the computers 
available at the time of this work.  Consequently, we carried out a significant
programming effort to produce a parallel implemention of VH-1, the details of
which were a significant part of a doctoral dissertation \citep{parr98}.
The parallel implementation and code testing are also described in the dissertation.

\section{Simulations of Stellar Mass Loss in a Hot Galactic Medium}

\begin{figure}
\plotone{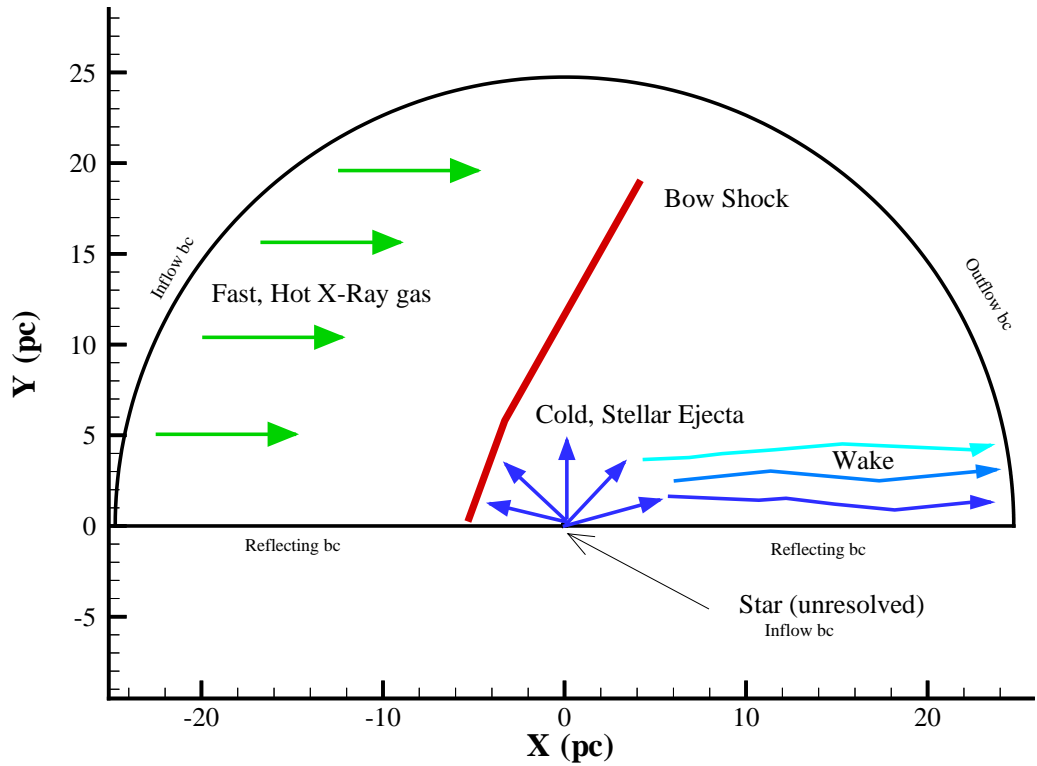}
\caption[Cartoon of the simulation]%
        {Cartoon of the simulation.  The hot ambient medium flows from left
	to right across the grid past the fixed star, encounters the cold 
	stellar ejecta and a bow shock is formed.  Momentum transfer pushes 
	the ejecta into a wake behind the star that is full of internal shocks 
	and vortices.  The contact surface between the ejecta and ambient
	medium is unstable to Kelvin-Helmholtz instabilities.  The grid boundary
	conditions are also shown.}
\label{fig:cartoon}
\end{figure}

There are certain common properties of all of the simulations, which
involve a hot, fast ambient medium gas 
flowing past a mass-losing star that is fixed at the origin of the
axisymmetric grid. The hot, fast ambient medium gas flows from left
to right across the grid with velocity $v_{amb} = \sigma_{\star}$, density 
$n_{amb}$, and temperature $T_{amb}$.  The pressure is determined from the ideal
gas law.  The outer boundaries have fixed inflow procedures on the left and 
fixed outflow procedures on the right with a check to make sure that the outflow
pressure does not exceed the inflow pressure.  The inner boundary represents a 
surface beyond the actual stellar surface with fixed inflow procedures.  The 
gas inflow from this boundary is the cold, slowly moving stellar wind, with a
strictly outward radial velocity of $v_{ejecta}$, density, $n_{ejecta}$, and 
temperature $T_{ejecta}$.  The mass loss rate that this corresponds to will be
a constant value for all runs, and for all time.  The exception to this is
a non-steady state simulation of a planetary nebula-like mass loss 
episode (to be discussed elsewhere). The axial grid boundaries implement a simple
reflecting boundary condition.  A cartoon of this scenario, after a bow shock
and wake have formed, is shown in Figure~\ref{fig:cartoon}.  Every simulation
is run from a physical time of 0 years to $3 \times 10^{6}$ years.  The 
calculated wake flow time of $\sim 8 \times 10^{5}$ years, so we
conservatively allow one million years for the flow to become 
quasi-steady-state.  Any steady-state averaging is done for simulation
times between one and three million years.  There are 10 runs:  five in the
2-D parameter space of ($v_{amb}$, $n_{amb}$), and five more at the 
same $v_{amb}$ and $n_{amb}$ with radiative cooling turned on.

\subsection{The Fiducial Simulation}

We find it useful to define a reference or fiducial simulation, against which
other simulations are compared.  
The parameters used for the fiducial simulation represent average physical
quantities near the effective radius for a typical intermediate luminosity elliptical 
galaxy.  This first simulation does not include the effects of radiative cooling.  
We will examine in detail the important physical processes that
drive the heating and cooling of the stellar ejecta, given a particular
set of input parameters.  Once we have established our analysis for this
fiducial run, it will serve as a reference point for examining the influence
of varying the ambient conditions, and radiative cooling.
The parameters for this (FDFV) and subsequent simulations are shown in 
Table 1.  Each simulation is run with a single metallicity
radiative cooling routine turned off and on.
A particular model is designated by a series of four or five letters, where F stands
for the fiducial values, so for the range of density, fiducial density 
conditions are designated FD, high density regions HD, and low density
regions as LD.  This is followed by two letters describing the flow velocity,
where the fiducial velocity is FV, high velocity is HV, and low velocity is LV.  
The detailed densities and velocities associated with these runs are given in Tables 1, 2.
Simulations that also have cooling are designated with a C as the last character.
So a simulation with the higher density, with the fiducial velocity and with radiative
cooling is designated HDFVC.

In this fiducial simulation, the hot gas velocity is $v_{amb} = 350$ km s$^{-1}$, which corresponds to a
one-dimensional velocity dispersion $\sigma_{\star} \approx 200$ km s$^{-1}$ from
the relationship $v = \sqrt{3}\,\sigma$.   This velocity dispersion is 
representative of a low to intermediate luminosity galaxy (\citealt{cft87};
\citealt{mkatlm97}).  The density of the gas is $n_{amb} = 10^{-3}\,{\rm cm}^{-3}$, 
which is a typical value for the galaxy.  Since we have used values
typical of intermediate $L_{X}$ galaxies, we set the temperature of the
X-ray gas $T_{amb} = 3 \times 10^{6}\,{\rm K} \approx 0.3\,{\rm keV}$.
These parameters lead to a gas moving slightly supersonically, with a
Mach number $M \approx 1.4$ for $\gamma = 5/3$, and a pressure 
$n T = 3 \times 10^{3}\,{\rm K cm}^{-3}$.  This pressure is
typical of the gas in the middle to outer regions of an elliptical galaxy.

The stellar mass loss rate that we use here 
($\dot{M}_{star} \approx 10^{-7} M_{\odot} yr^{-1}$), and for the other 
simulations, is a time-weighted average value for the red giant and 
asymptotic giant branch stages of stellar evolution for a solar-type star
\citep{r75,fbr90}.  The input parameters for
the ejecta inflow boundary that result in this rate are $v_{ejecta} = 35$ km s$^{-1}$,
and $n_{ejecta} = 1\,{\rm cm}^{-3}$, with a temperature 
$T_{ejecta} = 10^{4}\,{\rm K}$.  These are all average quantities for a
stellar wind at $r = 0.1$ pc.  A simple mass flux estimate at 
$r = 0.1\,{\rm pc}$ gives $\dot{M}_{star} \approx 4.8 \times 10^{-8} M_{\odot} yr^{-1}$.
This mass loss rate is half the rate that we want from the
inner boundary, but given these input values we obtain an actual calculated 
mass flux near $\dot{M}_{star} \approx 10^{-7} M_{\odot} yr^{-1}$.  This is due to high
pressure gradients near the boundary that affect the inflow.

\begin{figure}
\plotone{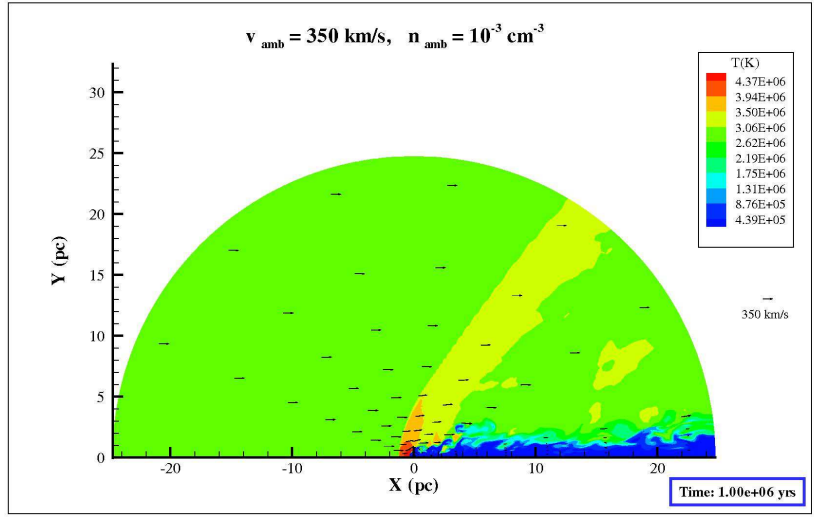}
\caption[Temperature map of the fully developed quasi-steady state flow]%
        {Temperature map of the fully developed quasi-steady state flow in 
	pressure equilibrium with the surrounding medium at a simulation time of
	$10^6$ years.  This is the fiducial run without cooling, FDFV. The bow shock and unstable 
	wake are fully formed.  Numerous Kelvin-Helmholtz instabilities are 
	formed along the contact surface of the wake.  The velocity field 
	is shown by the arrows, with the magnitude set by the reference vector 
	at the right.}
\label{fig:full}
\end{figure}

As material begins to flow into the ambient medium from the boundary
representing the stellar outflow, a contact discontinuity
and bow shock are formed.  The ambient flow
imparts momentum to the stellar ejecta that pushes it into a downstream wake
that is in pressure equilibrium with the ambient gas.  
There are a few fluid instabilities that occur in this situation.
When a shock wave impulsively accelerates a density interface, small
amplitude perturbations grown and, in the non-linear regime, one fluid
component penetrates the other.  The conditions appropriate for this 
Richtmyer-Meshkov instability appear to occur at the forward contact
surface of the bow shock, a situation common for all simulations.
Further downstream in the flow, there is a cool wake with a relatively
rapidly flowing hotter and lower density medium.  This situation is 
unstable to the growth of Kelvin-Helmholtz instabilities.  
As these instabilities travel down the wake, vortices are formed, further
mixing the two gases.  In addition, cool material that extends or detaches
from the wake or head region can collide and shock with the fast low-density
material flowing by.  Within these interactions, the conditions for Rayleigh-
Taylor instabilities are sometimes established, and this instability also helps
to mix the wake gas with the ambient material.
Because of these processes, the amount of cold ($T \leq 10^4$ K) gas is greatly 
reduced by the time the wake exits the grid in this non-radiative cooling run.
The entire grid for this simulation after a quasi-steady state flow has been
established with a fully formed bow shock and complex wake is shown in 
Figure~\ref{fig:full}.

A bow shock is formed upstream from the stellar inflow boundary since the 
ambient gas is mildly supersonic.  The standoff distance between the inner 
boundary and the vertex of the bow shock and contact discontinuity is consistent with 
theoretical expectations.  The standoff distance from a mass losing star is found
by balancing the ram pressures of the ambient medium and stellar wind,
$\rho_{amb} v_{amb}^2 = \rho_{ej} v_{ej}^2$, where 
$\rho_{amb},\;v_{amb},\:\rho_{ej},\;{\rm and}\;v_{ej}$ are the
mass density and velocity of the ambient medium and ejecta at the contact, 
respectively.  
Solving for the standoff distance from the center of the star $r_0$ given 
that $\rho_{ej} \propto r^{-2}$, yields \citep{bkk71,dyson75,wilkin96}:

\beq
\label{eq:standoff}
r_0 = \left( \frac{\dot{M}_{star}\,v_{ej}}{4 \pi\,\rho_{amb}\,v_{amb}^2} \right)^{1/2}\:.
\eeq

\begin{figure}
\epsscale{0.75}
\plotone{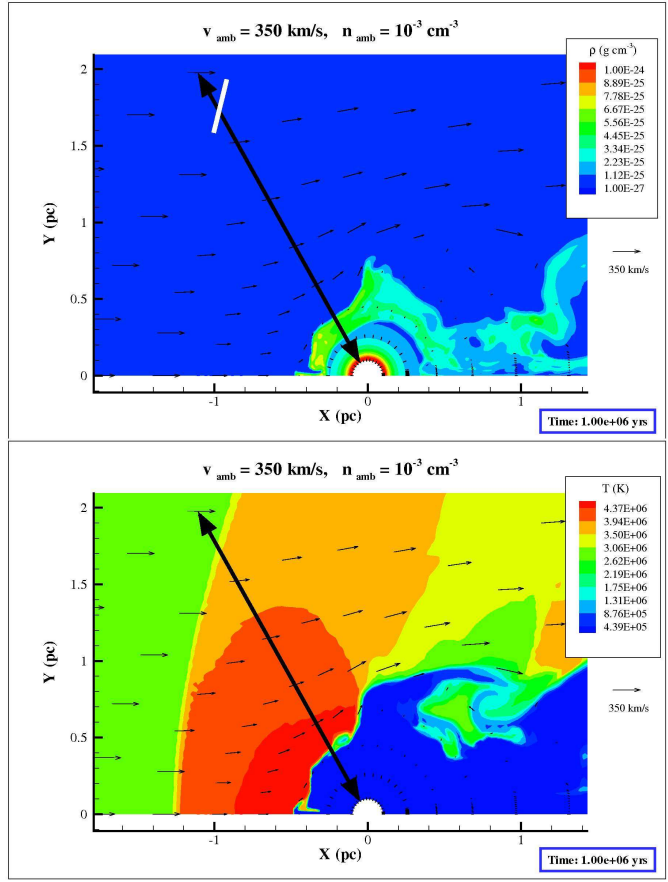}
\caption[Contour maps of the leading edge collision]%
        {Temperature maps of the leading edge collision between the stellar wind
	and the ambient flow, with density shown in the top panel, and 
	temperature in the bottom panel (for the fiducial run FDFV).  The black arrow shows the line where
	data were extracted to make the plots in Figure~\ref{fig:khheaddat}.
	The maps show the bow shock and contact discontinuity, as well as a
	developed KH instability that is peeling off to produce a larger
	effective area for the ambient medium to shock heat.  The small leakage
	along the upstream reflecting symmetry axis that is caused by numerical
	error is also shown.  The white
	line on the density map marks the location of the bow shock.  The
	velocity field is shown by the smaller black arrows.}
\label{fig:khdtheadl}
\end{figure}

\noindent The simulation typically places the contact discontinuity near 
$r_0 \approx 0.4$ pc (see Figure~\ref{fig:khdtheadl}), which is almost precisely
the calculated value of $r_0 \approx 0.4$ pc, given input values of
$\dot{M}_{star} = 10^{-7} M_{\odot} yr^{-1},\:v_{amb} = 350\:{\rm km\,s}^{-1}, 
\rho_{amb} = 2 \times 10^{-27}\,{\rm g\,cm}^{-3},\;
v_{ej} = 35\:{\rm km\,s}^{-1}$ which are for a boundary at $r = 0.1$ pc.  The 
precise location of the shocks varies
with time as large instabilities peel off the contact and present an ejecta
region with a temporarily larger effective cross-section.

\begin{figure}
\epsscale{1.0}
\plotone{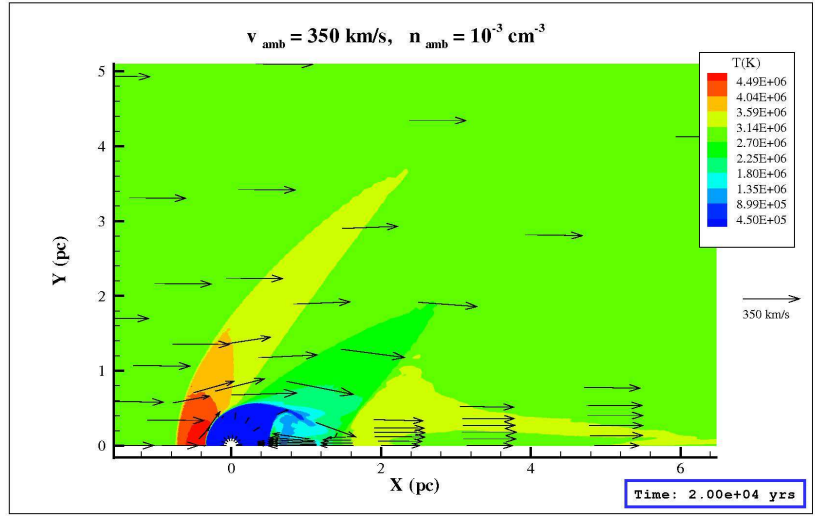}
\caption[Temperature contours showing the initial bow shock formation]%
        {The temperature map showing the initial bow shock formation in
	the $M \approx 1.4$ flow, due to the inflow from the inner boundary (run FDFV).
	A large eddy formed on the immediate downstream side of the outflow
	can also be seen.  The velocity field is shown by the arrows, with
	the magnitude set by the reference vector at the right.}
\label{fig:initbow}
\end{figure}

The main function of the bow shock is to thermalize the ambient gas 
as the ambient flow imparts momentum to the ejecta, pushing it into a wake.  
The early formation of this bow shock is shown in Figure~\ref{fig:initbow}.  
Shock heating from the bow shock and contact discontinuity is the 
mechanism invoked by the standard theory to locally thermalize $100\%$ of the 
stellar material to the velocity dispersion temperature.  
However, the simulation makes it clear that these shocks are not heating all the 
ejecta, and do not appear to be the main heating mechanism.

\subsection{The Role of Kelvin-Helmholtz Instabilities}

The relative tangential velocity difference between the stellar mass loss and
the ambient medium causes Kelvin-Helmholtz (hereafter KH) instabilities to
develop.  In the linear regime the growth rate of the KH instability 
$\sigma_{KH}$, with the wave number $k = 2 \pi / \lambda$, is given by 
(e.g., \citealt{lamb45}):

\beq
\label{eq:khrate}
\sigma_{KH} = k \left[ \frac{\rho_{amb}\,\rho_{ej}}{(\rho_{amb} + \rho_{ej})^2}
(v_{amb} - v_{ej})^2 \right]^{1/2}\:,
\eeq

\noindent where $\rho_{amb},\;v_{amb},\:\rho_{ej},\;{\rm and}\;v_{ej}$ are now
the mass density and {\em tangential} velocity of the shocked ambient 
medium and ejecta, respectively.  
The e-folding growth time $\tau_{KH} \simeq 1/\sigma_{KH}$ of
modes for the present case with $\rho_{amb} \ll \rho_{ej}$ and 
$v_{amb} \gg v_{ej}$ is approximately given by:

\beq
\label{eq:khapptime}
\tau_{KH} \approx \left( \frac{\rho_{ej}}{\rho_{amb}} \right)^{1/2}\:
\frac{\lambda}{2 \pi v_{amb}}\:,
\eeq

\noindent where $\lambda$ is the length scale of growing mode.

In considering the growth of KH modes, we note that modes with $\lambda$ below the numerical 
resolution of the grid and algorithm are suppressed.  The grid
resolution at the contact surface at the head of the flow is 
$\Delta r \approx \Delta \theta \approx 0.01$ parsecs, while the elongated cells 
half the distance down the high-resolution wake region are
$\Delta r \approx 0.1$ parsecs, $\Delta \theta \approx 0.01$ parsecs.  These
modes might be important in the evolution of the wake material, but we cannot 
address them.  We will calculate e-folding growth times for typical
conditions present in the contact discontinuity and wake, where KH instabilities
are seen to form and grow.  These times will be shown to typically be 100 times
shorter than the flow time (i.e. the time for the wake to be advected off the
grid).  Since the flow is quasi-steady state after roughly $8 \times 10^5$ years,
we assume that any results from times later than this are typical of the 
flow evolution.  The numbers used for the current analysis were taken from
the grid state at a simulation time of $10^6$ years.

\begin{figure}
\epsscale{0.8}
\plotone{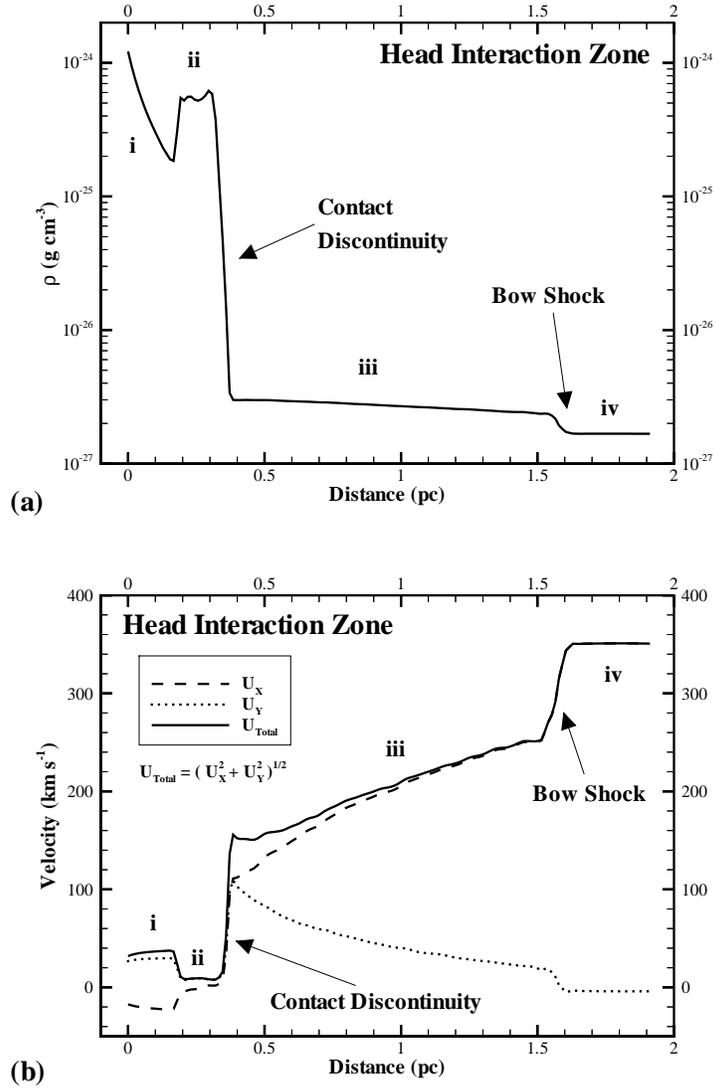}
\caption[Selected profiles of the leading shock zone]%
	{Selected profiles of the leading shock zone from data taken along the 
	line designated in Figures~\ref{fig:khdtheadl}, run FDFV.
	(a) Mass density profile, with the contact discontinuity and bow
	shock designated.  The abscissa gives the distance in parsecs along
	the line, beginning at the inner grid boundary.  Region specifications: 
	i) the undisturbed stellar wind, $r^{-2}$ adiabatic expansion; ii) 
	post-shock region from the contact discontinuity shock; iii) post-shock 
	zone from the bow shock; iv) unaffected $M = 1.4$ hot ambient medium.
	(b) Velocity profiles, with the same specifications as the mass density
	profile.  The X-direction, Y-direction, and combined velocities are
	shown.  The extraction line was chosen to cross the contact 
	discontinuity at a place where the ``ambient'' and wind velocity vectors
	were aligned (see text),  so ${\rm U}_{{\rm Total}}$ was used for the 
	velocities in the KH growth time calculations.}
\label{fig:khheaddat}
\end{figure}

We will first examine the conditions for the contact discontinuity at the
leading edge of the ejecta-ambient collision.  Figure~\ref{fig:khdtheadl} shows
mass density and temperature contours for the region including the vertex of 
the bow shock,
the main contact discontinuity, and the first growing KH instabilities.  The
temperature map also shows the upstream leakage of a small amount of cold gas 
along the reflecting symmetry axis due to unavoidable numerical errors (see 
above).  Since we are interested in the steady state
wake flow, the influence of this axis error should be minimal.  The
bow shock is not visible on the density map since the density drop across it 
is very small, but its location is indicated by the short white line.  The
black arrow represents the line where the data were extracted to make the
density and velocity plots in Figure~\ref{fig:khheaddat}.  
The data were extracted from this line because
the velocities ambient gas and the ejecta are almost totally parallel to each
other and tangential to the discontinuity at this point, which is a
necessary condition for the velocities in Equation~\ref{eq:khapptime}.
After substituting what appears to be a typical base length for some beginning
instabilities as the length scale $\lambda \sim 0.2$ pc, and the other
relevant values ($\rho_{amb} \sim 3 \times 10^{-27}\,{\rm g\,cm}^{-3},\;
\rho_{ej} \sim 5 \times 10^{-25} {\rm g\,cm}^{-3},\;{\rm and}\;
v_{amb} \sim 155\:{\rm km\,s}^{-1}$) into the linear 
Equation~\ref{eq:khapptime}, we obtain $\tau_{KH} \approx 3000$ years for 
the leading edge flow.  It should be noted that the ``ambient'' values used here 
were actually the post-shock gas values since that is the fluid actually at 
the contact discontinuity (inner shock).  The flow time is simply found
by dividing the radius of the grid by the speed of the wake $v_{ej}$, so that
$\tau_{flow,wake} \approx 8 \times 10^5$ years.  As stated above, $\tau_{KH} \ll
\tau_{flow,wake}$, so there is ample time for these instabilities to form.
It should be noted that the modes that form do so with a very high 
degree of spatial and temporal accuracy, given an inner cell size of order 
$0.01$ pc on a side, and a Courant time step that is typically 
$t_{Courant} \approx 5$ years.  The longer modes will take longer to form
and we see that the larger instabilities present themselves further
downstream, and these less frequent modes appear to do a great deal of wake 
mixing.

\begin{figure}
\epsscale{0.75}
\plotone{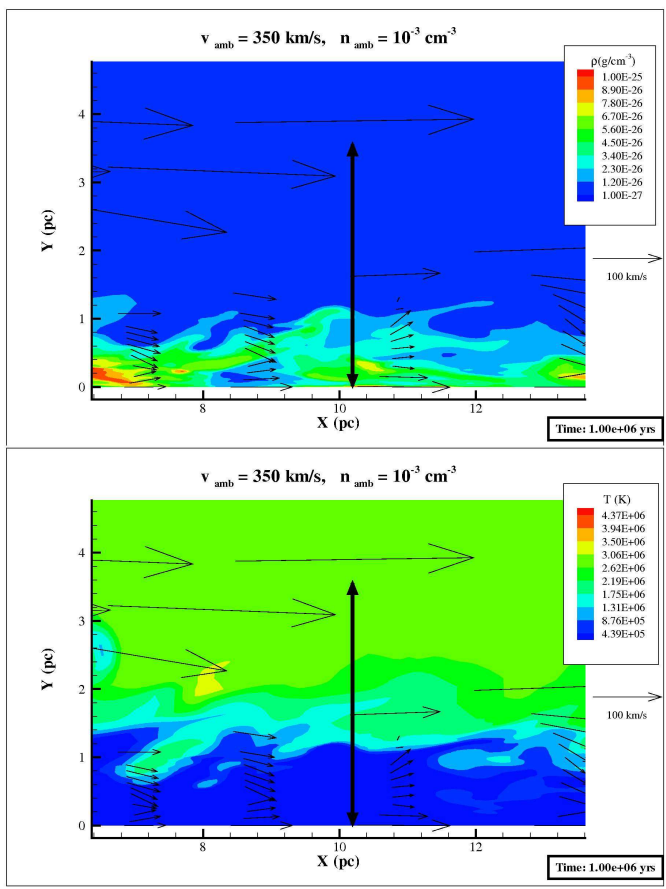}
\caption[Contour maps of a typical wake region]%
        {Temperature maps of a typical wake region, with density shown in the top 
	panel, and
        temperature in the bottom panel (run FDFV).  The vertical black arrow shows the line where
        data were extracted to make the plots in Figure~\ref{fig:khwakedat}.
	The discontinuity interface here is a much wider shear layer, but
	KH instabilities still grow.
        The maps show a very unstable and complex wake with numerous KH
	``fingers'' and accompanying vortices being advected along the wake.
	The velocity field, shown by the smaller black arrows, shows the
	mixing taking place, as alternating areas of
	high and low density (pressure) between KH eddys.}
\label{fig:khdtwake}
\end{figure}

\begin{figure}
\epsscale{0.75}
\plotone{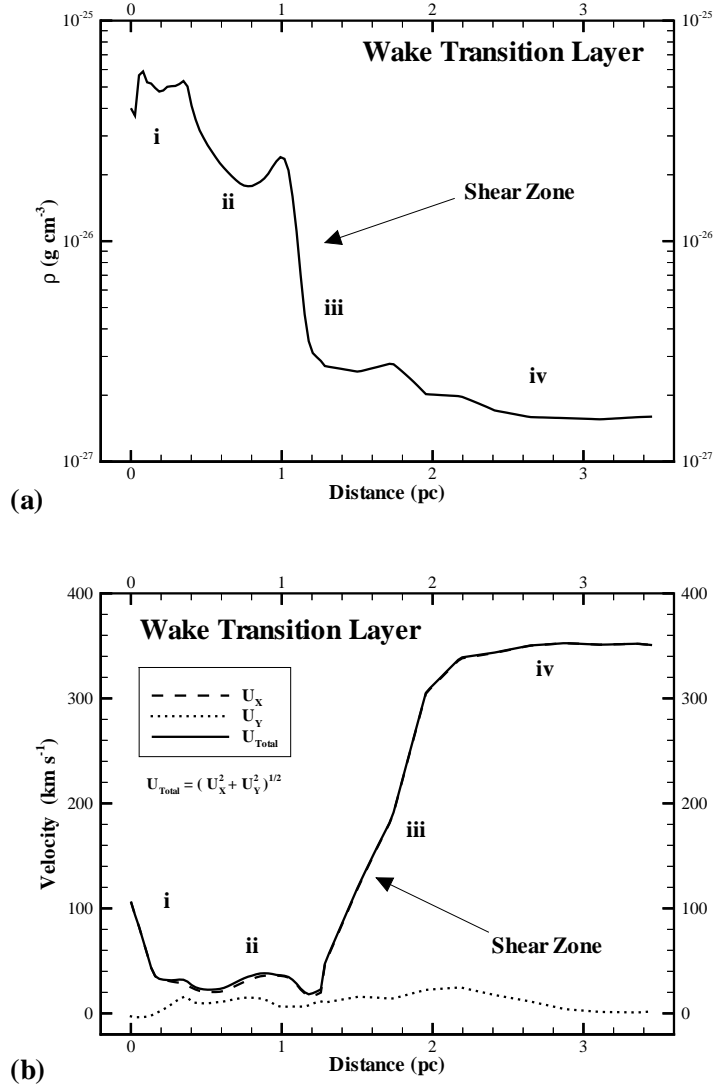}
\caption[Selected profiles of the leading shock zone]%
        {Selected profiles of the leading shock zone from data taken along the
        line designated in Figure~\ref{fig:khdtwake} (run FDFV).
        (a) Mass density profile, with the wide shear zone designated.
 	The abscissa gives the vertical distance in parsecs, beginning at the 
	bottom edge of the grid.  Region specifications:  i) more dense and
	fast wake material, which could be partially numerical in origin;
	ii) the main wake area filled with instabilities and vortices; iii) the
	wide mixing shear zone between the wake and hot medium; iv) relatively
	unaffected ambient material, with the exception of an occasional
	disruption by large advected KH ``finger.''}
\label{fig:khwakedat}
\end{figure}

The second KH instability measurements are from a typical region in the 
downstream wake from at the same simulation time.  Instead of the clear 
two-shock structure seen in the head region, the wake consists of the
cold ejecta material with significant mixing taking place through the many KH 
instabilities and
the vortices that they form.  This unstable colder material is in pressure
equilibrium with the ambient medium, but it is still separated from the
hot gas by a more broad discontinuity that we will call a shear layer.
Figure~\ref{fig:khdtwake} shows mass density and temperature contours for this 
region of the wake, as well as the line where data were extracted.  As stated 
above, since the flow is quasi-steady state, we can assume that the place where 
we have extracted data to be average and typical for the wake flow.  The mass
density and velocity profiles are shown in Figure~\ref{fig:khwakedat}. 
The data used in the KH time calculations are taken on each size of the full
shear layer.  Since the wake and ambient flows are almost entirely in the 
X-direction, with the exception of vortices, the velocity in that direction 
is appropriate for the KH calculations.  The density contrast between
the wake and the hot gas is not as large as at the head contact discontinuity,
so we need to use Equation~\ref{eq:khrate} to find KH growth times rather 
than using Equation~\ref{eq:khapptime}.
Substituting the relevant wake parameters for the ejecta values 
($\rho_{amb} \sim 2 \times 10^{-27}\,{\rm g\,cm}^{-3},\;
\rho_{wake} \sim 3 \times 10^{-26} {\rm g\,cm}^{-3},\;{\rm and}\;
v_{amb} \sim 300\:{\rm km\,s}^{-1}\;,
v_{wake} \sim 30\:{\rm km\,s}^{-1}$), and with a longer empirically determined
length scale of 
$\lambda \sim 1$ pc, we obtain $\tau_{KH} \approx 3000$ years for 
the wake region.  The longer length scales in the wake region compensate for
the smaller density contrasts compared to the head region, so that $\tau_{KH}$ 
is approximately the same for both regions.  This time scale is
over 100 times shorter than the flow time, so the instabilities have plenty
of time to grow before they are advected off the grid.  The long modes take
a longer time to form, and we see large instabilities in the wake that
we do not see in the head region, presumably because they do not have time to
form before being advected downstream.

These simulations show that Kelvin-Helmholtz (KH) instabilities have ample time and
resolution to grow in all areas of the ejecta-ambient gas interface.  The
instabilities, and the vortices that form directly downwind of them, are
the primary mechanisms for heating by mixing the cold ejecta into the hot 
ambient gas.  There also seems to be a characteristic scale for instabilities, 
but a full modal analysis would be necessary understand its origin.

\begin{figure}
\epsscale{1.0}
\plotone{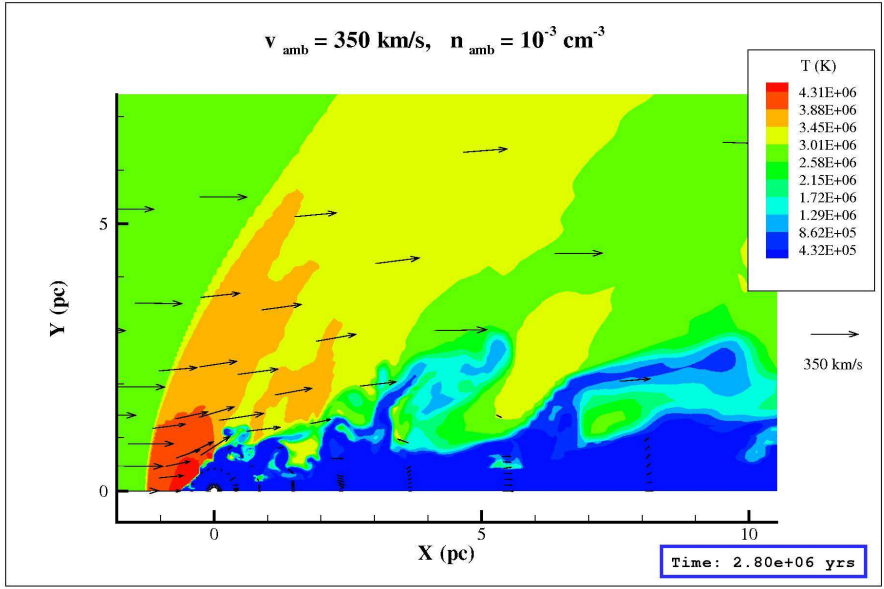}
\caption[Temperature map of growing KH ``fingers'']%
        {Temperature map of growing KH ``fingers'' that are advected downstream,
	and continuously shocked and bent by the ambient gas (run FDFV).  The largest modes
	can reach quite far into the surrounding gas, and even become detached
	from the wake altogether.}
\label{fig:tkhshocks}
\end{figure}

The Kelvin-Helmholtz (KH) instabilities discussed in the previous section
are the main cause of mixing in the wake.  If a large wavelength mode
grows under pressure equilibrium near the head of the flow, its presence 
creates a much larger effective
collision cross-section for the ambient medium.  The ambient medium imparts
momentum to these large KH ``fingers,'' and very effectively heats them and
advects them downstream above the main wake zone.  Occasionally large pieces 
become completely detached from the wake, and mix even more with the 
ambient gas.
Figure~\ref{fig:tkhshocks} shows growing modes from the initial contact
discontinuity that form long fingers that are bent and advected downstream.
These colder pieces are continuously shocked by the ambient gas as they move
down the wake.

Although many smaller KH instabilities grow and advect down the flow, the 
mixing (heating) effects of the largest ``fingers'' are far greater, since 
they can dredge up more inner wake material.  We can calculate
the location in the wake where we expect these large modes to present themselves
by multiplying their growth time by the wake flow speed.  A KH mode with a
wavelength of five parsecs that begins growing near the head of the flow will
have an approximate growth time of $\sim 7.5 \times 10^4$ years.  Given an 
outer wake flow speed of $50$ km s$^{-1}$, this mode should be growing large by 
about four parsecs downstream, which is consistent with the simulations.

\begin{figure}
\epsscale{0.75}
\plotone{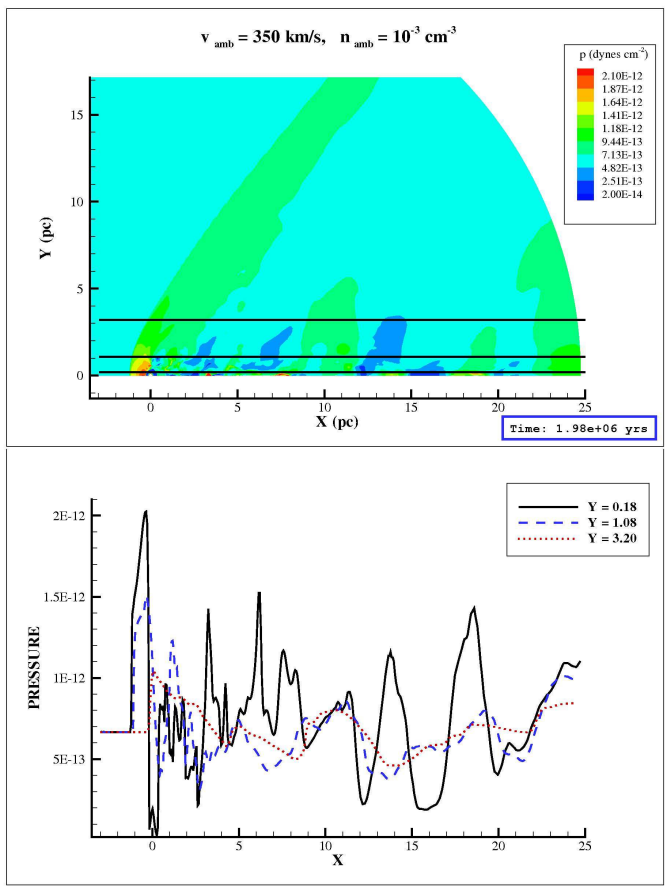}
\caption[Pressure map and profiles for the entire wake]%
	{Pressure map and profiles for the entire wake, with the horizontal black lines
	on the map designating the locations where data were extracted
	to make the profiles in the lower panel (run FDFV).  These profiles were taken
	at increasing distance away from the symmetry axis 
	($Y = 0.18,\;1.08,\;3.20$ pc), and show the complex nature of the wake
	compared to the region dominated by the ambient flow.  The common peaks
	seem to suggest characteristic ``wavelengths'' for the flow on the
	order of $\sim 5$ pc.  The dip in the $Y = 0.18$ profile occurs because 
	it passes through the pure
	stellar wind region near the inner inflow boundary.}
\label{fig:paltfull}
\end{figure}

Now that we have established that the location of these modes agrees with
predictions, we will examine their considerable influence on the dynamics of
the wake.  The large blobs and their accompanying vortices 
move down the wake creating alternating large zones of relatively low and high 
pressure in the wake.  This effect is shown by the pressure map in 
Figure~\ref{fig:paltfull}, which also designates three horizontal lines along
which pressure profiles were extracted and plotted.  These pressure profiles, 
taken at increasing distance from the symmetry axis, show the dramatic 
alternating changes in pressure that occur down the wake.  The pressure gradient
of some of the pressure jumps suggest significant shocks traveling downstream
{\em within} the wake in conjunction with these fingers.  A temperature
map of the same region would only show shocks occurring where the large fingers
continue to reach out in the ambient gas.  The common peaks in these curves
seem to suggest characteristic ``wavelengths'' for the flow on the order of
$\sim 5$ pc.  They are spreading apart as they travel downstream because the
wake experiences a slight acceleration downstream due to the constant influence
of the fast ambient flow.
The other main effect of the KH instabilities, regardless of wavelength,
comes from the vortices that always form downstream from a growing mode,
with the longer modes having larger-scale vortices.
These vortices help to create weak shocks as they 
mix ambient gas into the interior of the wake.  This suggests that they play a
more important role in heating the stellar ejecta than the initial bow shock
and contact discontinuity.

\subsection{Cumulative Heating}

The discussion of heating mechanisms has been necessarily quite qualitative,
but we can obtain an estimate for the cumulative heating by calculating the amount
of stellar mass loss that remains cold as a function of distance downstream in the 
wake.  We do this by computing the time-averaged mass flux, $4 \pi r^2 \rho v$, 
through shells of increasing radius for increasing temperature cutoffs.  
Time-averaging is possible once a quasi-steady state flow has
been established.  If the mass flux for the colder temperature cutoff bins 
decreases significantly (i.e. by one or two orders of magnitude) before reaching
the edge of the grid, then we will consider that gas as being heated.  Both this
overall decline in mass flux and the nature of the profile are important,
since the profile will tell us something of the mechanism primarily responsible
for heating that gas.  A slow steady decline in the amount of cool material suggests wake vortex and shock
heating as the gas is advected downstream.  A sharp decline near the head of 
the flow is due to a strong bow shock and the contact discontinuity heating upon
initial contact with the ambient medium.

We calculate these time-averaged mass fluxes during a 
post-processing step that takes grid data for a given time, computes
the fluxes as a function of radius, and then averages them with every other 
output between the time when the quasi-steady state flow is established
and the end of the simulation.  Since the wake flow time is roughly
$8 \times 10^5$ years, we conservatively use grid outputs for simulation
times between $10^6$ years and $3 \times 10^6$ years, when the simulation ends.
For each preset temperature cutoff, we step through increasing radii 
($0.1 - 25$ pc), and compute the combined mass flux for each angular cell i
($0 - \pi$ rad) at that radius with an average temperature at or below the 
current cutoff value.  If a cell has radial velocity that is negative 
(i.e. not flowing out through the shell), we do not include it in the sum.
The outward mass flux for a single cell is given by:

\beq
\label{eq:mfluxcalc}
\dot{M}(r,\theta) = 2 \pi r^2 \Delta\theta\:\rho \max(v_r, 0)\:, 
\eeq

\noindent where $\Delta\theta$ is the angular extent of the outer cell
boundary, $r = 0.1$ pc defines the inner boundary, and $\theta = 0$ corresponds
to the downstream symmetry axis.  Since we sum $\dot{M}(r,\theta)$ for 
$\theta = 0 - \pi$ radians at a given $r$, the total area of the shell
is the full $4 \pi r^2$ cm$^2$.  Also, since we only count gas flowing outward 
through these shells (via the $\max(v_r, 0)$ term), the only material included 
for $\theta > \pi/2$ is the stellar wind before it reaches the contact 
discontinuity around $r = 0.4$ pc.  Without this condition, the large negative
mass flux effects of the hot wind for $\theta > \pi/2$ would obscure all of 
the other mass flux data in which we are interested.  This condition on the 
cell velocity direction will not include material flowing backward due to 
vortices, but such an effect should be averaged out over time as the vortices 
are steadily advected downstream.

\begin{figure}
\epsscale{1.0}
\plotone{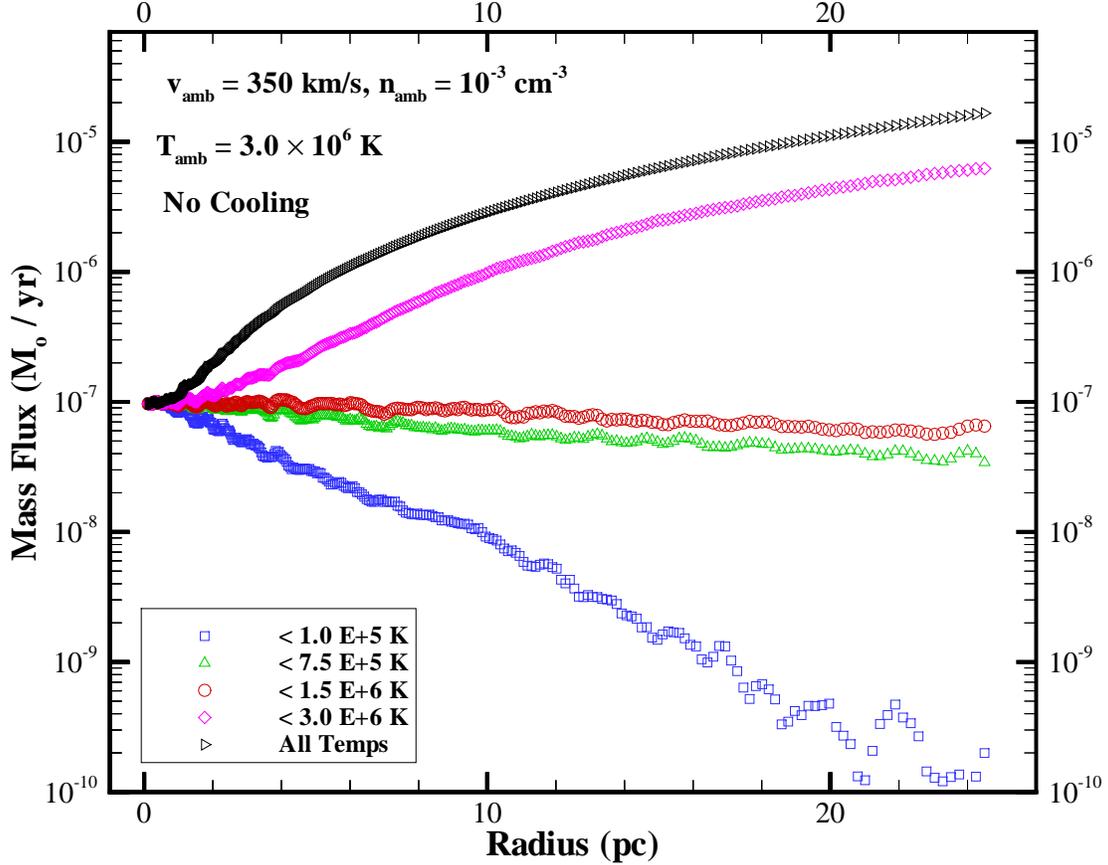}
\caption[Mass fluxes for the fiducial simulation without cooling]%
	{Mass fluxes for the fiducial simulation without cooling (run FDFV), as a function
	of radius.  The difference between the ``All Temps'' line and the
	$T_{amb} = 3 \times 10^6$ K line represents gas shock heated to above
	the ambient temperature.  As expected, the value near the inner boundary 
	equals the 
	stellar mass loss rate, which is constant for the pure wind region 
	corresponding to zone (i) in Figure~\ref{fig:khdtheadl}.  All of the
	profiles for temperature cutoffs at or below $T_{amb}/2$ experience a
	steadily decreasing mass flux, with the $10^5$ K profile steadily 
	dropping to a value roughly three orders of magnitude lower than 
	$\dot{M}_{star}$.  These profiles suggest constant heating of the cold 
	gas as it flows downstream.}
\label{fig:mffidnc}
\end{figure}

The mass flux profiles are shown in Figure~\ref{fig:mffidnc}, where the 
five cumulative temperature cutoffs are at
three increasing fractions of the ambient temperature $T_{amb}$, at $T_{amb}$,
and for all temperatures (to include shocked gas).  Each 
profile includes all the gas with a temperature at or below the cutoff value.
Heating of any gas below $T_{amb}$ is indicated by that profile falling
below the stellar inflow rate of $10^{-7} M_{\odot} yr^{-1}$.  A steep decline suggests
a high heating rate.
Because of the definition of the mass flux $\dot{M}(r)$, if there were no mass
loss from a star, $\dot{M}(r) \propto r^{-2}$ , and the ``All Temps'' line
would follow this form exactly.  The ``All Temps'' and $T_{amb}$ curves would be
identical.  In the simulations with stellar mass loss,
the ``All Temps'' curve quickly approaches the $\dot{M}(r) \propto r^{-2}$ form
as the ambient mass flux exceeds the mass flux from the star.

The mass flux near $r = 0.1$ pc is constant in the stellar wind regime and equal to the 
stellar mass loss rate $\dot{M}_{star} = 10^{-7} M_{\odot} yr^{-1}$.  Also, the gap between
the ``All Temps'' and $T_{amb} = 3 \times 10^6$ K line is caused by the 
gas that is shock heated above the ambient temperature.  The profile that
includes every temperature rises rapidly at first, suggesting that the initial 
bow shock and contact discontinuity are the cause.  Both this profile and
the $T_{amb} = 3 \times 10^6$ K profile continue to rise from this point 
because the hot ambient flow accounts for the bulk of the mass flux at larger
radii as the surface area of the shell increases as $r^2$.
However, it is the fate of the cooler stellar ejecta that is of greatest interest here.
If the stellar ejecta did not interact with the hot ambient medium, the stellar
ejecta mass flux would be a horizontal line at the contant value of 
$10^{-7} M_{\odot} yr^{-1}$ and with a temperature below $10^5$ K.

The new information provided by this analysis comes from the behavior of the
profiles for cumulative temperatures at and below half the ambient value.
Each of these profiles steadily falls below the initial stellar mass loss 
rate.  The mass flux in the lowest temperature bin, $T = 10^5$ K, decreases by three orders of
magnitude, suggesting significant heating of the coldest stellar ejecta before
it exits the grid.   The first result is found in the changing slope of the profiles as they 
steadily decrease with increasing distance downstream.  As mentioned above, 
such a steady decrease suggests that the gradual vortex mixing and shocking 
in the wake are responsible for most of the heating rather than the initial bow shock and 
contact discontinuity.  The other important point is 
that all of the cold gas should eventually be heated, and at a fairly 
significant rate given the slope of the profiles.

Now that we have discussed the nature of the interaction in the fiducial case
without cooling, and detailed our analysis of the heating processes, we will
examine simulations that include the effects of radiative cooling on the flow.

\section{The Fiducial Simulation with Radiative Cooling}

For a system in equilibrium, the cooling of the hot gas as a result of 
optically thin radiation at X-ray energies is the mechanism opposing 
the heating effects discussed above.  It is precisely this radiation that
lead to the initial discovery of the hot gas in elliptical galaxies.  We
are concerned with how the addition of radiative cooling will effect the
dynamics of interaction, and we would expect the key parameter to be
the ratio of the cooling time $\tau_{cool}$ to the time for the
wake to be heated and mixed in the absence of heating (we will refer to this
as the flow time of the ambient medium, $\tau_{flow,amb}$).  
These properties of the ambient medium will 
determine the properties of the cold gas since the wake tries to achieve pressure 
equilibrium with the surrounding medium.  If the cooling time is much longer
than the flow time, we would expect very little to change with the addition of 
the radiative loses.  In the discussion of the rest of the simulations, we 
will examine the nature of the interaction as we change the value of this
ratio.

Although the optically thin radiative cooling function $\Lambda_{N}(T)$ can be 
approximated by a power law over limited ranges in temperatures, here
we used the full calculated cooling function of \citet{sd93} for a solar
metallicity plasma.
The cooling function is dominated by collisional metal lines for the
temperature range appropriate for our calculations, and so the magnitude of the
function varies roughly linearly with metallicity:  $\Lambda_{N} \propto Z$.
The actual functional change of $\Lambda_{N}$ with metallicity is shown
in \citet{sd93}, where this roughly linear gradient can be clearly seen.

The key parameter for the simulations 
with radiative losses is the ratio of the cooling time to the flow time of the 
ambient gas.  Since the cooling rate per unit volume is $n^2 \Lambda_{N}(T)$,
we can change the cooling strength and thereby the cooling time by adjusting 
the density $n$, the temperature $T$, or the metallicity.  The wider range of expected
density and its higher power dependence makes density the more attractive 
parameter to vary, and so we will keep the gas metallicity constant by using only the solar 
cooling function.

In order to demonstrate that cooling should affect the dynamics of the flow,
we can calculate some typical cooling times for the ambient gas in pressure
equilibrium with the wake.  The instanteous isochoric cooling time for a 
fully ionized $90\%$ hydrogen, $10\%$ helium (by mass) gas is given by

\beq
\label{eq:tcool}
\tau_{cool} = \frac{4}{\gamma - 1}\:\frac{\mu m_{p} k_{b} T}{\Lambda_{N}(T) \rho}\:,
\eeq

\noindent where $\gamma$ is the ratio of specific heats, $\mu$ is the mean 
molecular weight, $m_{p}$ is the proton mass, $k_{b}$ is Boltzmann's constant, 
and $\rho$ is the mass density.  We will always use $\gamma = 5/3$ and
$\mu = 0.62$ in this work.
In order to calculate this cooling time for the flow, we use average values of the 
physical parameters for several regions of the flow.
As expected, the cooling time for the undisturbed tenuous hot ambient medium is 
very long, $\sim 10^9$ years.  However, the cooling times for the
post-shock gas in the head and wake of the flow are $\sim 6 - 10 \times 10^4$ 
years, which are less than the flow time of $\sim 3 \times 10^5$ years.
Therefore, we can expect cooling to be important since these times are of the 
same order.

\begin{figure}
\epsscale{0.75}
\plotone{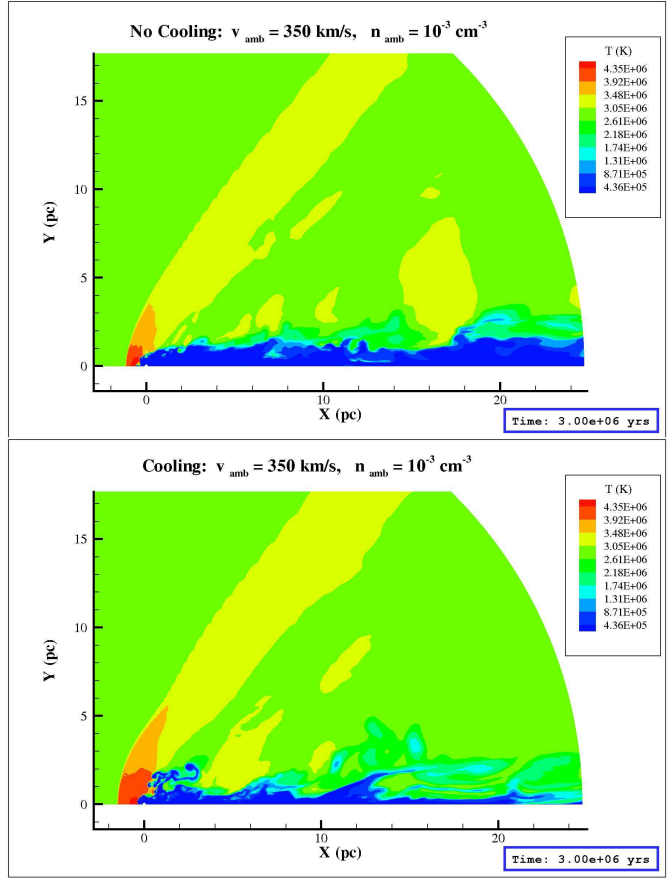}
\caption[Typical maps for cooling vs non-cooling fiducial runs]%
        {Typical maps for cooling vs non-cooling fiducial runs taken 
	at a simulation time of $3 \times 10^6$ years.  The simulation without
	cooling is in the top panel (run FDFV), and the 
	run with the addition of radiative losses is in the bottom panel (run FDFVC). These 
	temperature maps, using the same scales, show significant differences
	in the bow shock, in KH instability behavior, and in the concentration
	of the wake.}
\label{fig:fidcompfull}
\end{figure}

\begin{figure}
\epsscale{0.75}
\plotone{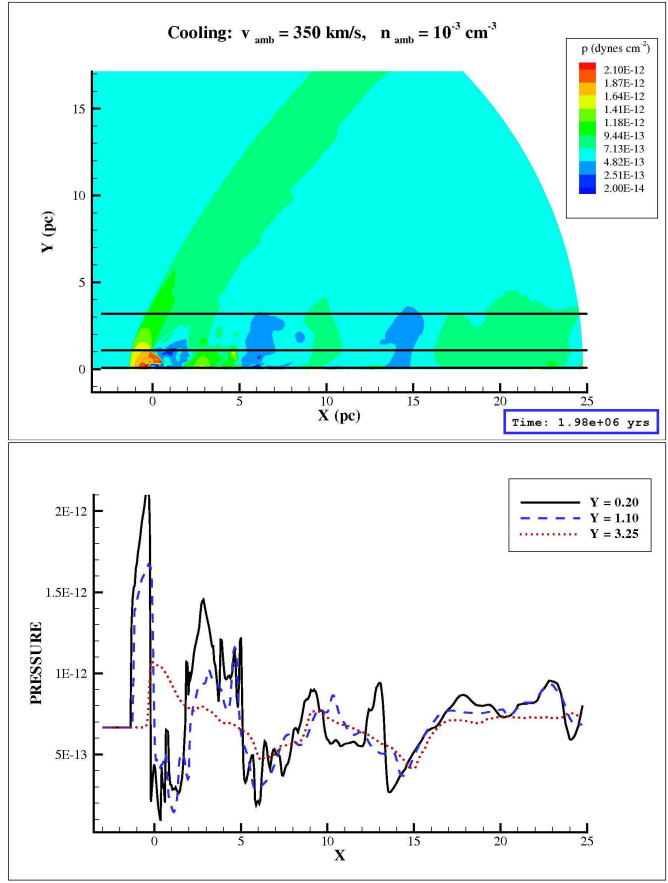}
\caption[Pressure map and profiles for the entire wake]%
        {Pressure map and profiles for the entire wake, with the black lines
        on the map designating the locations where data were extracted
        to make the profiles in the lower panel (run FDFVC).  As for the non-cooling case, these profiles were taken
        at increasing distance away from the symmetry axis
        ($Y = 0.20,\;1.10,\;3.25$ pc), and while the wake is less complex in
	this case, there are still large KH modes moving downstream that cause 
	the pressure fluctuations.}
\label{fig:paltfullcool}
\end{figure}

There are three obvious differences between the cooling (FDFVC) and non-cooling
(FDFV) simulations at the same simulation time
(e.g., Figure~\ref{fig:fidcompfull}):  the bow shock, the Kelvin-Helmholtz
instability behavior, and the wake entrainment.  The large mode instabilities
moving downstream are similar to those in the non-cooling case, and this is 
shown in Figure~\ref{fig:paltfullcool}.

\begin{figure}
\epsscale{1.0}
\plotone{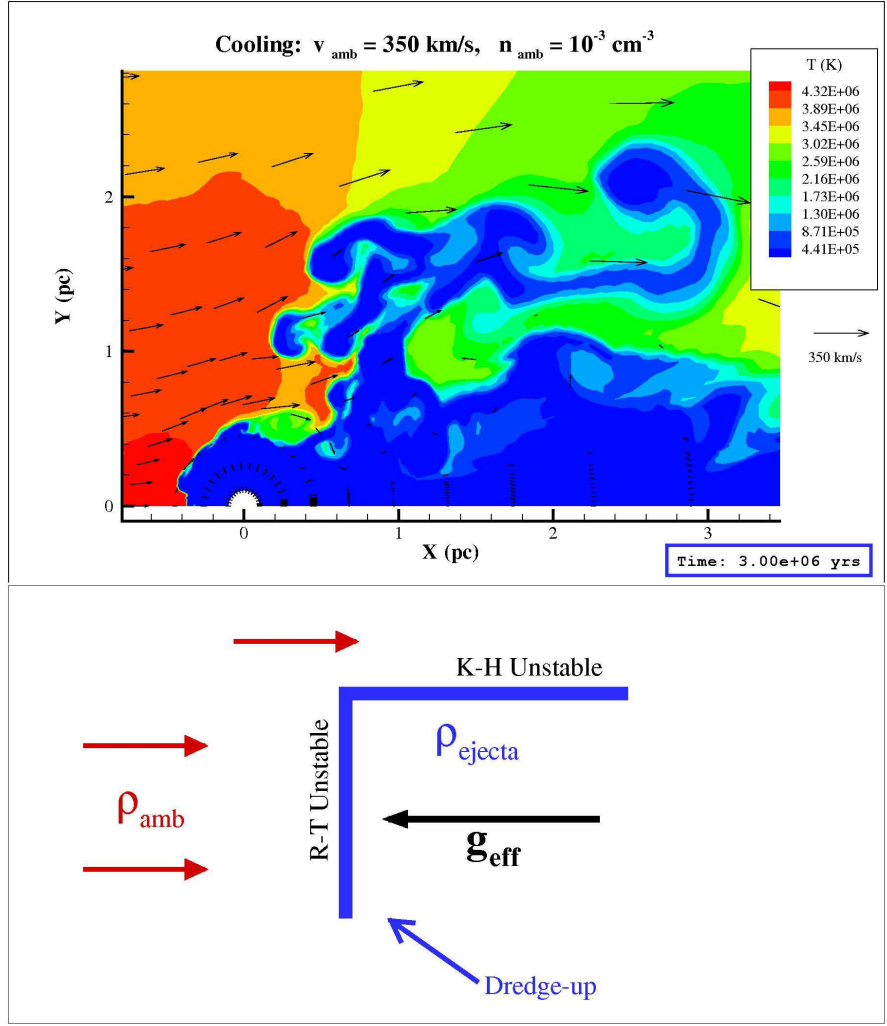}
\caption[Rayleigh-Taylor and Kelvin-Helmholtz instabilities]%
	{Rayleigh-Taylor and Kelvin-Helmholtz instabilities are both seen
	forming as this cold blob is dredged up and thrust into the ambient
	flow (run FDFVC).  The RT occurs because an effective gravitational force is
	set up in the opposite direction of the flow, and the KH occurs
	because the ambient flow is slipping past the cold blob.  The bottom
	panel shows a cartoon of this situation for a single blob.}
\label{fig:rtkh}
\end{figure}

First, the bow shock is stronger and more pronounced than the simulations
without cooling.  Although the contact 
discontinuity is in the same location for both cases, the cooling bow shock 
stands roughly $40\%$ further away from the inner boundary.  This is because
a large instability finger in the cooling case protrudes an entire 
parsec further into the ambient material than the in the non-cooling case. 
Therefore, the ejecta in the cooling case is presenting the ambient gas with an 
effective interaction cross-section area ($\pi R^2$) that is about four times
larger than the in non-cooling run.  This feature is not time-stationary, and
these gaseous fingers occur cyclically, developing because a large 
clockwise-rotating vortex immediately downstream from the head transports material
and pushes it out into the flow directly above the head.  
These fingers are quickly advected downstream and
broken up, but this vortex continues to dredge up cold wake gas.  The 
disruption of the cold finger generates small globules that have the distinct 
appearance of Rayleigh-Taylor instabilities (the familiar mushroom cap shape)
on the side finger facing the ambient wind.  
As the finger is bent to the right, it becomes unstable to Kelvin-Helmholtz 
instabilities as the cold gas moves parallel with the hot gas that is
flowing by (see Figure~\ref{fig:rtkh}).  The appearance of Rayleigh-Taylor 
instabilities is entirely reasonable because the deceleration of the hot gas as it encounters
the cold blob acts as an effective gravitational field in a direction normal to 
the interface.  In the frame of the cold blob, the effective gravitational force
is opposite in direction to the movement of the hot gas.  Therefore, the
classical Rayleigh-Taylor condition for instability is met because a dense gas
(the cold blob) lies above a tenuous gas (the hot ambient flow).  An illustration
of this is shown in the bottom panel of Figure~\ref{fig:rtkh}.
Upon averaging over time, simulations with cooling have a stronger and more pronounced bow shock.
A direct consequence is that more momentum is imparted to the ejecta material 
than in the non-cooling case.  Whereas Rayleigh-Taylor instabilities were
also identified in the non-cooling case, they are more common in this 
simulation with cooling.

\begin{figure}
\epsscale{0.75}
\plotone{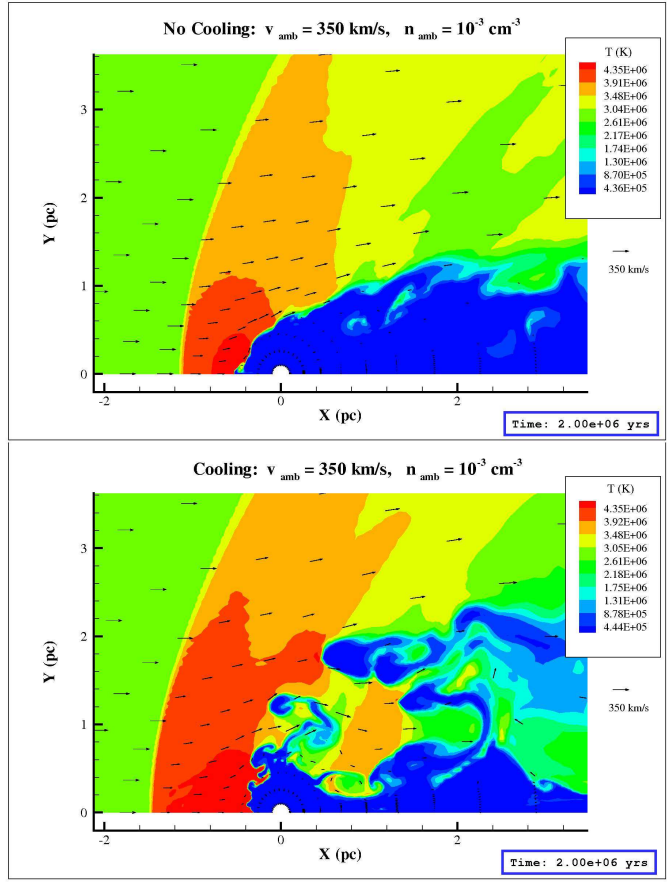}
\caption[Closeup maps showing the difference in KH instability growth]%
	{Closeup maps showing the difference in KH instability growth at a
	simulation time of $2 \times 10^6$ years (run FDFVC).  The difference in the
	general stability of the flow near the head of the interaction can be
	clearly seen in these temperature maps.  Cooling is shown in
	the bottom panel, with non-cooling in the top panel.  The velocity
	field is shown by the black arrows.}
\label{fig:fidcompclose}
\end{figure}

A second important difference between simulations is that the Kelvin-Helmholtz
instabilities are structurally different from those in the non-cooling case.
The interaction in the cooling case generally seems more chaotic and
unstable than in the non-cooling case.
The ``fingers'' that grow near the head of the flow in the cooling case 
typically grow more quickly and extend further into the ambient gas 
than in the non-cooling case. However, if we compare the KH growth time scales 
for the initial contact discontinuity in both cases, we do not find
any significant differences.  The cooling case does appear to
grow long mode instabilities while damping out most smaller modes
in the wake.  Another aspect is that the cooling has damped out the 
numerical artifact along the upstream axis that is seen in the non-cooling simulations.    
Both of these effects in the head region can be seen in Figure~\ref{fig:fidcompclose}.

\begin{figure}
\epsscale{0.75}
\plotone{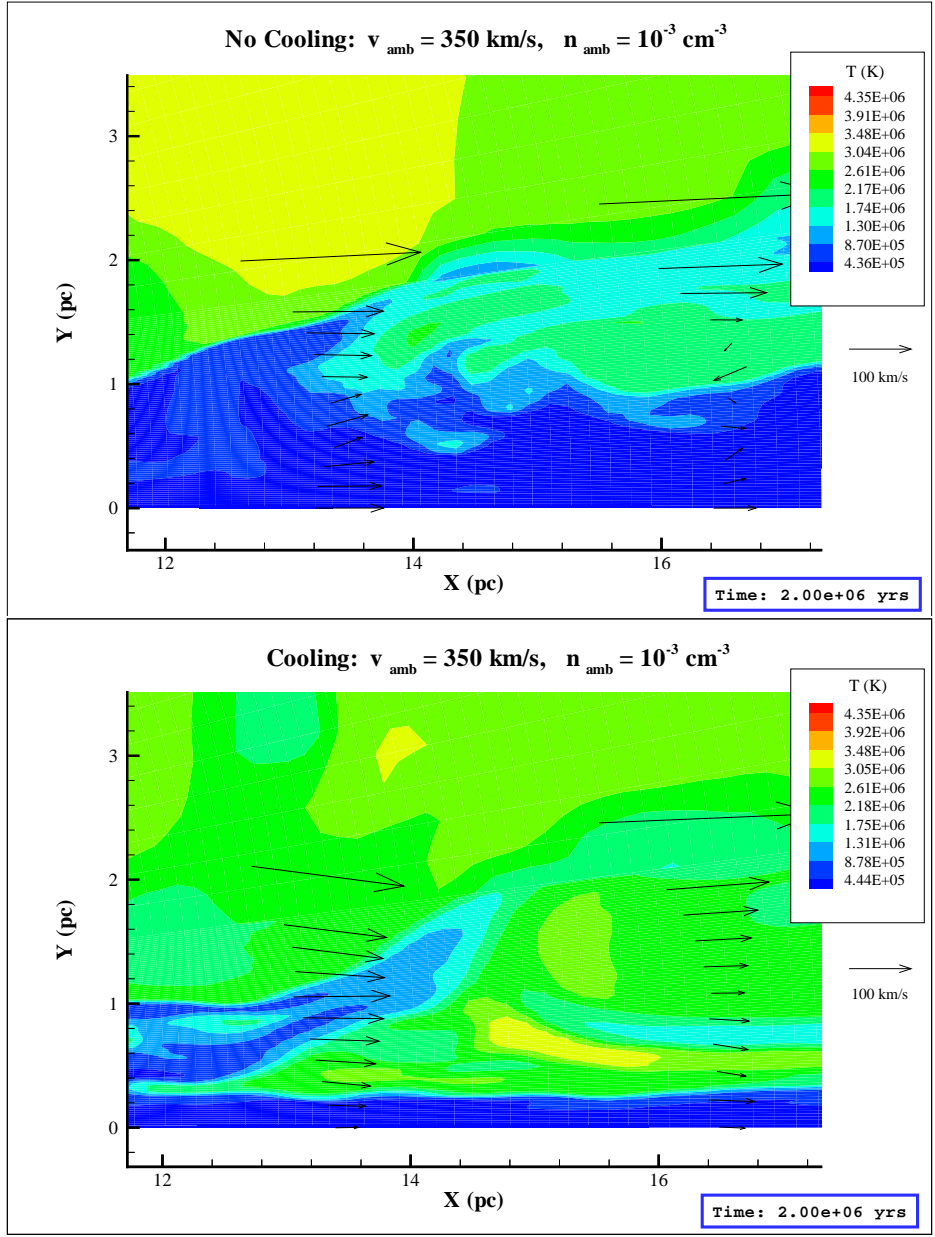}
\caption[Closeup temperature maps showing the difference in wake entrainment]%
        {Closeup temperature maps showing the difference in wake entrainment at a
        simulation time of $2 \times 10^6$ years.  The wake in the cooling 
	simulation (bottom panel; run FDFVC) is much more narrow and entrained than
	in the more unstable non-cooling case (top panel; run FDFV).  The velocity field 
	is shown by the black arrows.}
\label{fig:fidcompclosewake}
\end{figure}

The third important difference is that the wake is much more narrow 
in the cooling case (Figure~\ref{fig:fidcompfull}, Figure~\ref{fig:fidcompclosewake}).  
There is almost no vortex mixing going on in this part of the wake in 
the cooling case, with the exception of the widely spaced large KH modes
that survived from the initial contact region.  There are also fewer small
scale internal wake shocks generated as these larger modes move downstream.

\begin{figure}
\epsscale{1.0}
\plotone{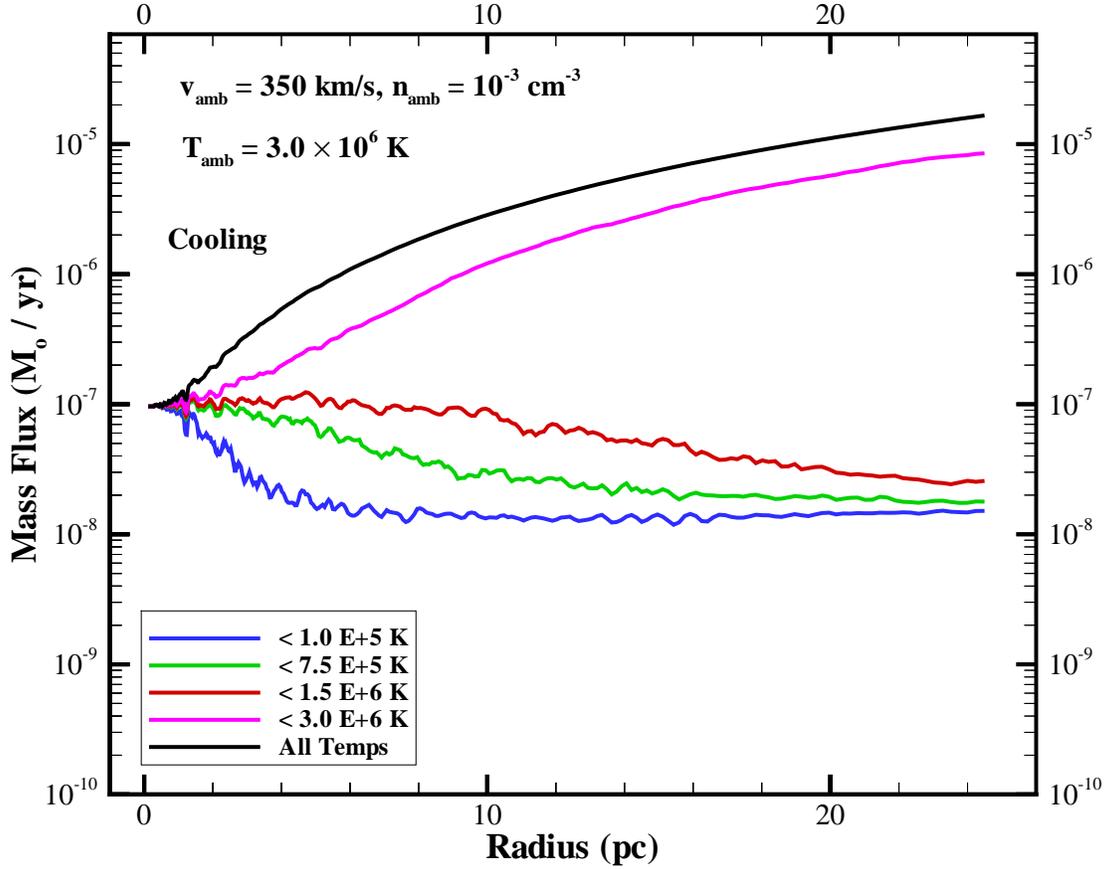}
\caption[Mass fluxes for the fiducial simulation with cooling]%
        {Mass fluxes for the fiducial simulation with cooling (run FDFVC), as a function
        of radius.  The data are arranged as in
	Figure~\ref{fig:mffidnc}.  The cold gas is promptly heated near the head of the flow in
	the first five parsecs, then remains at this level for the rest
	of the flow.  The cold gas fraction of the wake grows slightly 
	between 10 and 20 pc.  These profiles suggest heating of the cold gas by the
	bow shock and instabilities before it flows downstream.
	Once the cold gas settles into the more narrow wake, further net
	heating is effectively supressed by the radiative cooling.}
\label{fig:mffidc}
\end{figure}

Finally, there is a significant difference in the amount of mass that cools.
We calculate the mass flux in different temperature ranges, as 
discussed above (Figure~\ref{fig:mffidc}). The simulations show that most 
of the heating of the cold stellar ejecta takes place within the first few 
parsecs of the interaction.  The more unstable Richtmyer-Meshkov and Kelvin-
Helmholtz growth near the head of the flow 
leads to a greater momentum transfer and shocking from the ambient medium,
and prompt mixing and heating of the gas.  In the wake, radiative cooling causes
the damping of all but the largest modes that survive the trip downstream.  
The wake is largely devoid of small scale vortex mixing and shocking.   
Cold material that survives the initial contact region resides
in the very narrow and stable inner wake.
The fractional mass flux that remains cold is greater than in the non-cooling
simulation, and that fractional amount is increasing as it approaches the
outflow boundary of the simulation, presumably due to continued radiative cooling.
At the outflow boundary, the fractional amount of gas cooler than 10$^5$ K is 
about 10$^{-3}$ for the non-cooling case, but it is 0.15 when radiative cooling
is included.  Evidently, radiative cooling leads to a significant (but not dominant) 
fraction of the stellar ejecta remaining cool.

\section{The Influence of Ambient Gas Density}
The ambient gas density used for the fiducial simulations was chosen to be the
typical value in an elliptical galaxy.  However, the ambient gas density varies
by over an order of magnitude from the center to outer part, so here we 
investigate the consequences for the evolution of the stellar ejecta when the
density is an order of magnitude lower and half an order of magnitude greater than 
the values used in the fiducial runs.

\subsection{High Ambient Density Simulation}
\label{ssec-HDFV}

We consider the effects of increasing the ambient density to a value 
that is a reasonable average for the inner 5 kpc region of a typical galaxy
(e.g., \citealt{vtc88,dfj91}), where 
$n_{amb} = 3.33 \times 10^{-3}\,{\rm cm}^{-3}$ (see Table 1; simulation HDFV).
Because the stellar wake is in pressure equilibrium with the surrounding
medium, increasing the ambient density by 3.33 increases the pressure of the entire
system since both the wake and ambient gas temperatures have not changed.

\begin{figure}
\epsscale{0.75}
\plotone{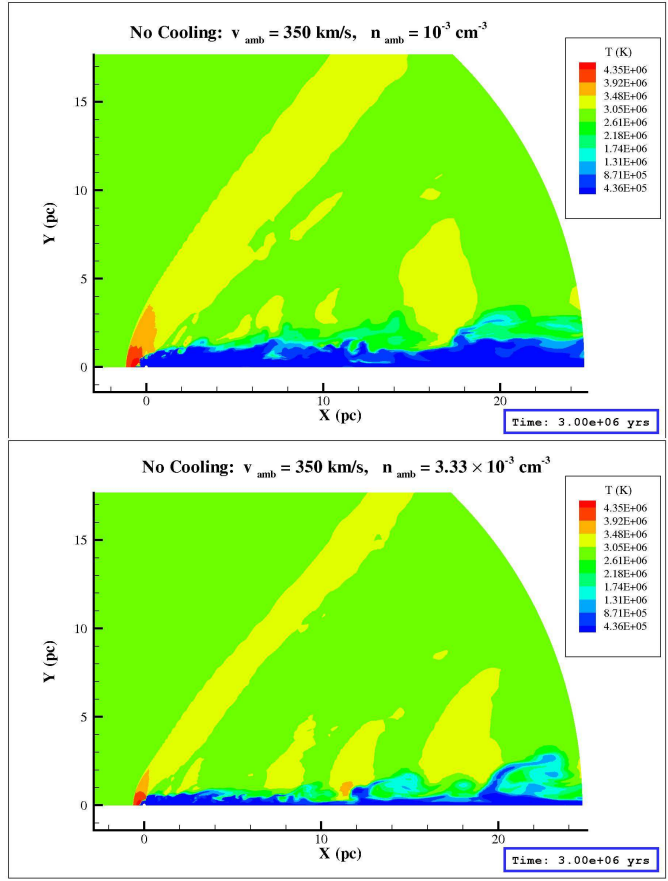}
\caption[Temperature maps showing the influence of higher ambient density]%
        {Temperature maps showing the influence of higher ambient density, at a
        simulation time of $3 \times 10^6$ years.  The flow in the higher density 
        simulation (bottom panel; run HDFV) occupies a smaller effective cross-section
	due to a higher ambient pressure,
	but the wake is as complex as in the fiducial (FDFV) simulation (top panel).}
\label{fig:hdfvfull}
\end{figure}

The higher ambient pressure present in this simulation (HDFV) causes the
wake to be more narrow than in the fiducial simulation (FDFV).  If we assume that 
the wake temperature is constant for both simulations, the size of the
wake should be reduced by a factor of $3.33^{-1/3} \approx 0.7$ when the
density is increased by a factor of 3.33 over the fiducial density:
$n_{wake} \propto h_{wake}^{-3} \propto p_{system},\;{\rm if}\; T = const$.
In agreement with this approximate scaling relationship, we find that the width
of the wake is reduced by roughly $\sim 60\%$ compared to the FDFV.  The initial contact 
discontinuity is also closer to the inner boundary because the relatively 
stronger ram pressure of the ambient gas balances that of the stellar ejecta at 
a smaller standoff distance $r_0$.  Also, this 
leads to a less pronounced bow shock.  A typical set of grid temperature maps
comparing this simulation with the fiducial one are shown in Figure~\ref{fig:hdfvfull}.  

While the wake may occupy a smaller volume in the high density simulation, it is
equally complex as in the fiducial run (Figure~\ref{fig:hdfvfull}).  
There are many large and small mode Kelvin-Helmholtz instabilities in the wake, 
along with the vortices and shocks that accompany them.  
This output time shows two large KH fingers and continuing shocks 
from the ambient gas that survived to large distances downstream.  
The theoretical KH growth time-scale is $\tau_{KH} \approx 3 - 4 \times 10^3$ yrs. 
The flow time $\tau_{flow}$ continues to be two orders of magnitude longer, so
we expect the flow to be unstable to the KH instabilities as it clearly the case.  

We would expect the heating of the gas to be relatively significant in
this simulation despite the reduced collision cross-section of the
flow, because the wake is highly KH unstable.  The internal mixing 
and shocking make the wake very complex, more so than in the fiducial simulation.
In the downstream direction, the cold stellar mass loss is heated more
efficiently than in the fiducial case.  The mass fraction of gas cooler than $10^5$ K
falls to three orders of magnitude below $\dot{M}_{star}$ in roughly 
half the downstream distance when compared to the fiducial simulation.
This is probably due partly to the higher thermal energy content of the
ambient medium.

The addition of radiative cooling (simulation HDFVC) has a very strong effect on the flow, 
especially on the fraction of cool gas in the downstream region.
Although the initial heating occurs more rapidly than in the fiducial case, the increase in the
pressure leads to a yet larger increase in the radiative cooling rate.
Because the volume cooling rate depends on the square of the density, this simulation should
have a cooling rate about ten times higher ($(3.33)^2$) than in the fiducial case.
The instantaneous cooling time for a typical region near the wake is 
$\tau_{cool} \sim 2 - 5 \times 10^4$ years, which is about 2 - 6 times less
than the fiducial density cooling time, and about half the flow time.

\begin{figure}
\epsscale{1.0}
\plotone{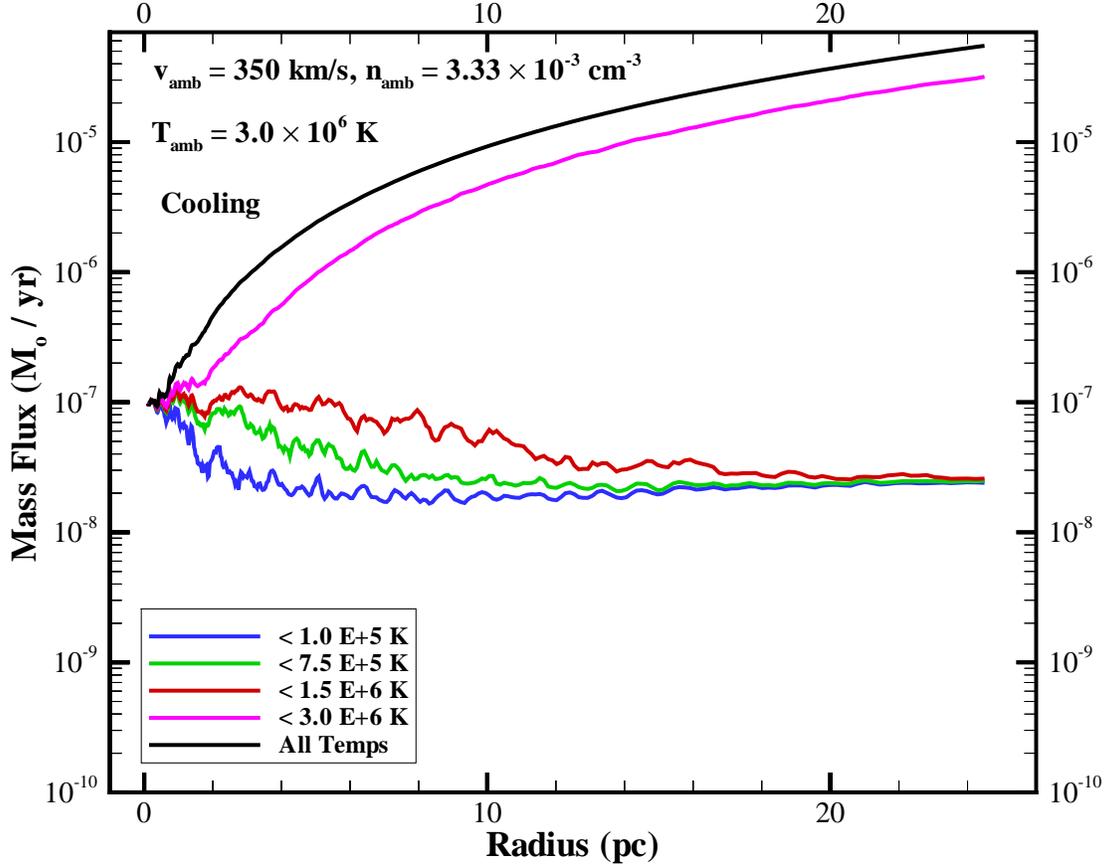}
\caption[Mass fluxes for the high density simulation with cooling]%
        {Mass fluxes for the high density simulation with cooling (run HDFVC), as a
        function of radius.  Most of the
        profiles for temperature cutoffs at or below $T_{amb}/2$ quickly 
	approach an asymptotic mass flux level roughly five times smaller than
	the stellar input value.  The bulk of the heating that is 
	accomplished by the strong bow shock and instabilities near the head
	of the flow.  Any colder gas that survives from this region 
	remains cool because of radiative cooling.}
\label{fig:mfhdfvc}
\end{figure}

As in the non-radiative case, most of the heating occurs quickly near the 
head of the flow.
Also, the narrowing of the wake is similar to the fiducial case when radiative
cooling was introduced (FDFVC).
Proceeding downstream, one finds a striking separation between temperature states
of the mass loss (Figure~\ref{fig:mfhdfvc}).  Gas cooler than $1.5 \times 10^6$ K
has cooled below $10^5$ K so that it has effectively become cold gas and will not
become part of the hot ambient medium.  This cooled gas is about one-quarter of the
mass lost from the star, and as this value does not change significantly through
the last 10 pc of the wake, it appears to be well-defined.

\subsection{Low Ambient Density Simulation}
\label{ssec-LDFV}

In this set of simulations (LDFV and LDFVC), the ambient density is reduced by factor of ten to
$n_{amb} = 10^{-4}\,{\rm cm}^{-3}$, which is representative of the outer regions 
of an elliptical galaxy. 
The lower pressure and momentum flux causes the contact discontinuity standoff 
distance to be about a factor of four larger than in the fiducial run.
The ram pressure of the ambient flow has been reduced by
an order of magnitude, and the bow shock is correspondingly less effective at 
heating and transferring momentum to the stellar ejecta.  
The RM, KH and RT instabilities do not develop rapidly enough to dredge 
up the inner parts of the expanded wake.
Consequently, relatively more heating occurs in the wake region.

For the case without radiative cooling (LDFV), the wake continues to be unstable to KH modes,
and RT fingers reach far into the ambient medium.  The lower system
pressure allows the cold wake material to fill a much larger volume,
and it does not appear that there are any KH modes of large enough amplitude to 
dredge up the inner wake along the axis.  So despite surface instabilities and
occasional gas parcels that break off from the wake, the central part of the 
downstream wake is less disturbed.  
The absence of instabilities large enough to dredge up the inner wake
suggest that they do not have enough time to grow because the 
preferred scale for the instability wavelengths and the growth times 
have been increased because of the lower density of the simulation.
Relative to the fiducial run, a larger fraction (0.11) of the gas cooler than $10^5$ K
remains when the wake exits the grid.  Since the mass fraction at this temperature
has not reached an asymptotic value, we expect that it would continue to decrease
had the grid been larger.

\begin{figure}
\plotone{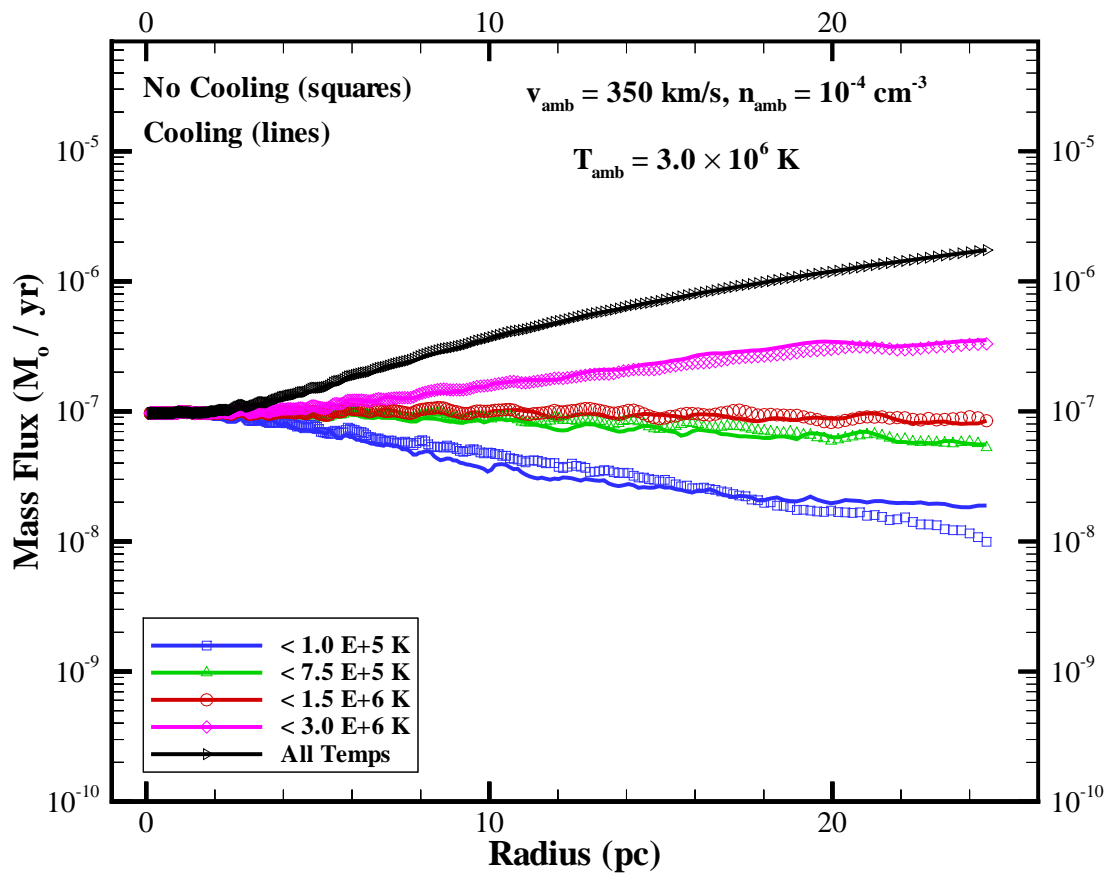}
\caption[LDFV mass flux with and without cooling]%
        {The mass flux with and without cooling for the low density simulations,
         LDFV and LDFVC.  The long cooling
	times due to the low ambient density and system pressure lead to 
	ineffectual cooling, so the curves should be quite similar.  
        However, the gas below 10$^5$ K appears to be approaching
        an steady value of about one-fifth of the initial stellar mass loss rate.}
\label{fig:mfldfvc-nc}
\end{figure}

When radiative cooling is added (LDFVC), the cooling times are $\tau_{cool} \sim 0.3 - 3 
\times 10^6$ years, depending on the location in the flow.
The geometric mean of the cooling time is about 10 times longer than in 
the fiducial case and about 10 times longer than the flow time.
Consequently, the influence on the dynamics of the flow are not great and the
width of the wake is similar in the simulation with and without radiative cooling.
Since the relative heating and mixing rate is relatively greater than the 
radiative cooling rate, we expected more of the stellar mass loss to be converted
into the hot ambient medium.
However, the mass flux of the gas cooler than $10^5$ K is reaching an asmyptotic
value in the final 5 pc of the flow at 0.2 of the initial mass 
loss rate(Figure~\ref{fig:mfldfvc-nc}).
In the last half of the flow, the cooling time in the cool wake is less than the flow time.

\section{The Influence of the Star's Velocity}
\label{sec-starvel}

Another important parameter that we consider is the velocity of the star
relative to the ambient medium.  To study this effect, we consider stellar 
velocities that are larger and smaller than the fiducial velocity by 125 km s$^{-1}$.
We anticipated that changes in the relative velocity would alter the heating
and cooling properties in the sense that the higher velocity simulations would
produce hotter shocked gas that has a significantly longer cooling time compared to the fiducial
case.  The reverse was expected for the lower velocity case, which also has an 
important difference from the other runs in that the stellar velocity is subsonic
compared to the sound speed of the ambient material.
Compared to our expectations, the simulations led to some surprising results
in that the efficiency of gas heating was different than initial expectations.
This is because a fraction of the gas is driven into a wake and the heating of
the wake occurs through the excitation KH instabilities of preferred wavelengths,
which appear to depend on the flow properties.  
As the heating rate of the cool wake depends on flow conditions, simulations are
the only approach to quantifying the efficiency with which stellar mass loss is
converted into hot ambient gas.
We simulate the different stellar space velocities by choosing different ambient 
gas velocities.  We continue to use the fiducial values of
the ambient density and temperature, as well as same stellar mass loss rate.

\subsection{High Stellar Velocity Simulation}
\label{ssec-FDHV}

In this simulation (FDHV), the Mach number is $M \approx 1.8$, compared to the value 
in the fiducial run of 1.3.  This leads to a higher ram pressure, a smaller 
standoff distance for the contact discontinuity, and a much more oblique bow shock 
that is closer to the contact discontinuity.

\begin{figure}
\epsscale{0.75}
\plotone{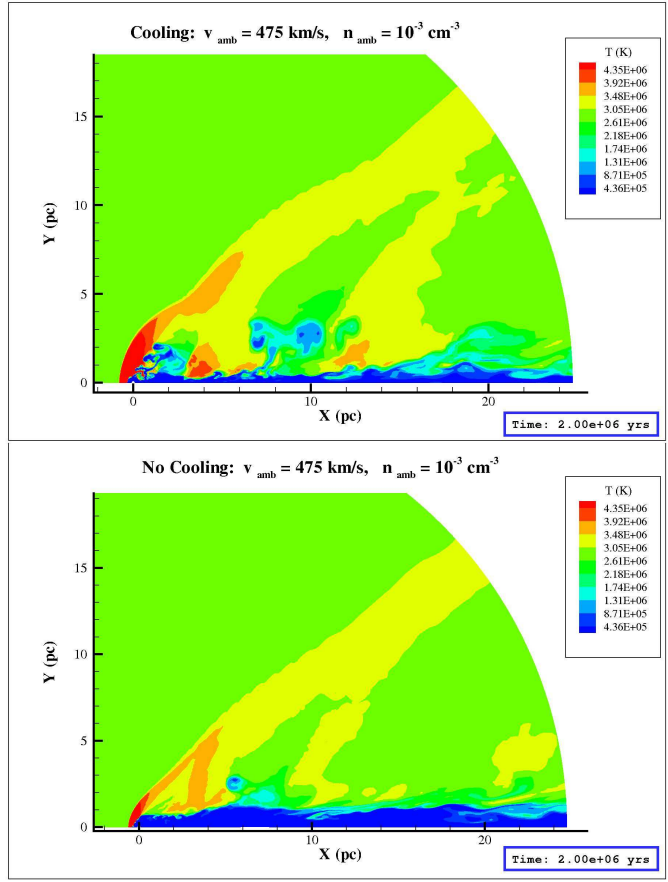}
\caption[Temperature maps showing the influence of cooling on FDHV]%
        {Temperature maps showing the influence of cooling on simulations with
        the fiducial density but high velocity (FDHV and FDHVC), at a
        simulation time of $2 \times 10^6$ years.   The cooled flow (top panel)
	is more unstable than the adiabatic wake (bottom panel).  Although the
	cooled wake is more narow as expected, the narrower
	large instabilities in the cooled case are also able to 
	survive longer against the fast wind.  This in turns increases mixing.}
\label{fig:hvcoolcomp}
\end{figure}

In the non-radiative simulation (FDHV), the KH modes begin to grow into filaments but they
become disrupted by the faster shocked ambient materia, so the filaments
do not grow as large as in the fiducial run.
The wake occupies roughly the same cross-section as in the fiducial case and it 
also contains instabilities, vortices, and internal shocks, but the faster flow 
prevents the growth of the large finger structures seen in the previous adiabatic 
wakes.  
The flow is unstable to Rayleigh-Taylor instabilities, but few structures
develop due to the more rapid disruption of the KH filaments.
In turn, this prevents large-scale mixing of the inner wake by these 
modes.  Most of the mixing occurs through relatively smaller modes.  
Another difference in the wake flow near the head is that a secondary shock typically forms
roughly four parsecs downstream (Figure~\ref{fig:hvcoolcomp}).
Due to the difference in the excitation of KH modes, one would expect 
a reduction in the cumulative heating rate of this non-cooling fast flow 
simulation relative to the fiducial case.  That expectation is verified by
the simulation, but the heating rate is still sufficient to reduced the 
mass flux of gas cooler than 10$^5$ K by two orders of magnitude at the point
where it exits the grid.

\begin{figure}
\epsscale{1.0}
\plotone{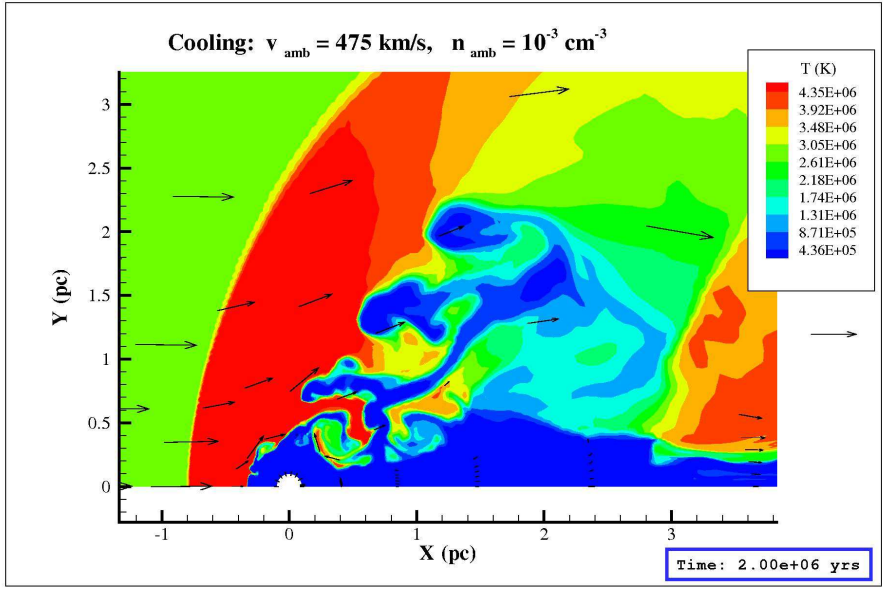}
\caption[Cooled Rayleigh-Taylor fingers in the fastwind]%
	{Cooled Rayleigh-Taylor fingers in the higher flow velocity simulation
	(FDHVC), at a simulation time of $2 \times 10^6$ years.  
	Condensed cooled blobs have become extended 
	and become separated due to the flow of the fast low-density material.  These classical
	Rayleigh-Taylor fingers and the prompt mixing that accompanies them
	are very similar to those seen in Figure~\ref{fig:rtkh}.}
\label{fig:hvrtc}
\end{figure}

The inclusion of radiative cooling (run FDHVC) makes several dramatic differences in the
results, especially in the unstable modes that are excited.  
Instabilities produce cool fingers near the head of the wake, which are 
excited and grow more rapidly than in the non-cooling case (Figure~\ref{fig:hvrtc}).
Further downstream, the wake is also more unstable and larger modes develop
before interacting with the fast surrounding flow (Figure~\ref{fig:hvcoolcomp}).
This change appears to be related to the width of the wake, which is narrower due 
to the loss of thermal energy through radiative losses.  In a more slender wake, the 
size of the most rapidly growing modes has changed, where larger and cooler fingers 
develop.

\begin{figure}
\plotone{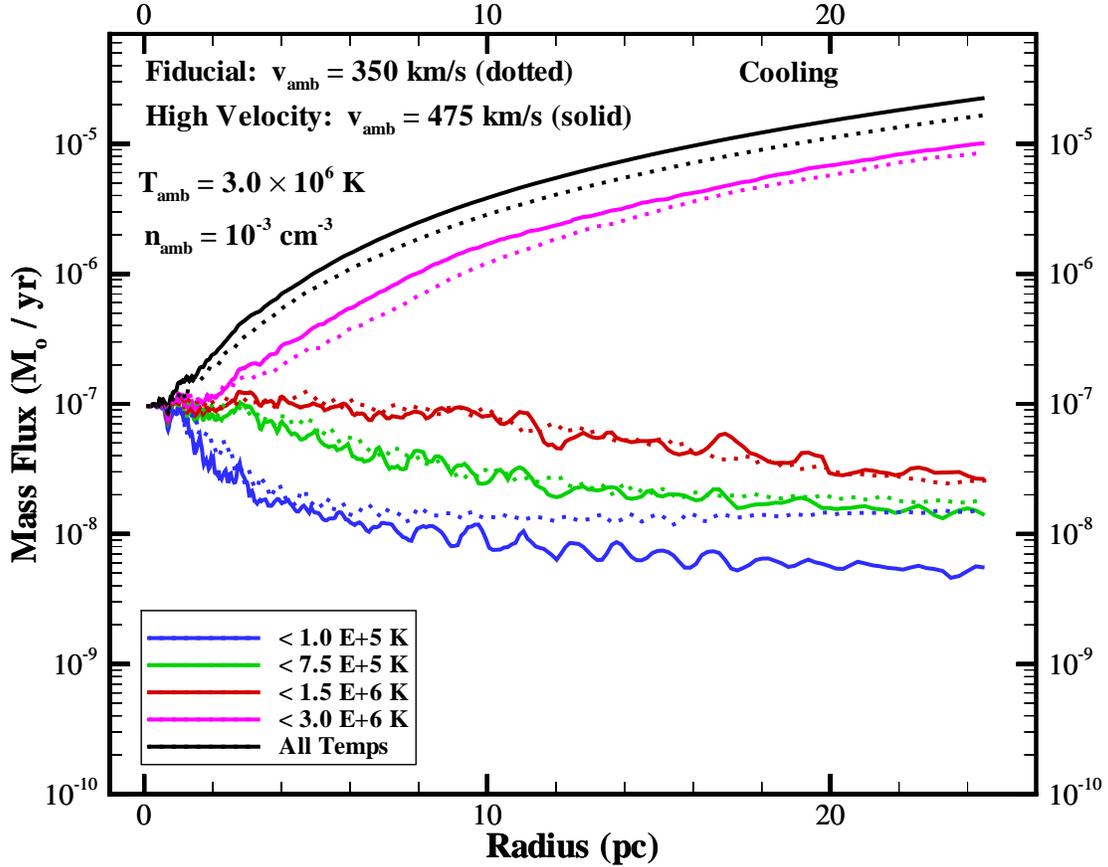}
\caption[Mass flux for cooled FDHV and FDFV]%
	{Mass flux for the fiducial and high flow velocity simulations with 
	radiative cooling (FDHVC and FDFVC).  For gas with a
	temperature below $3 \times 10^5$ K, the high velocity profiles (solid)
	lie below the fiducial velocity ones (dotted), suggesting
	the shock cooling time of the higher velocity case is indeed
	longer, and so the cooling is less efficient at lower temperatures.  
	This difference in efficiency does not seem to apply for higher 
	temperature gas.  Also, the lack of an asymptotic
	convergence for the colder FDHVC profiles indicates that the accretion
	of warmer gas onto the narrow central cooled wake is occurring
	more slowly, indicating less efficient cooling.}
\label{fig:mffdhv-fvc}
\end{figure}

As with the other flows that include radiative cooling, most of the heating occurs 
near the head of the flow.  This heating is hastened by the more easily excited 
instabilities, which undergo mixing and shocks.  The greater relative flow velocity
of the star leads to slightly more heating within the first few parsecs of the star
(relative to the fiducial run), but further down the wake, the higher velocity 
leads to a larger conversion of cold gas to hot gas (Figure~\ref{fig:mffdhv-fvc}).  
As the gas exits the grid, the remaining fraction of gas below 10$^5$ K is 
0.057 and the value is still decreasing.
The higher velocities produce higher temperature gas that cools more slowly, relative
to the fiducial case.
The effect of a higher stellar velocity is a more efficient conversion of the 
stellar mass lost into the hot ambient medium by at least a factor of three.

\subsection{Low Stellar Velocity Simulation}
\label{ssec-FDLV}

\begin{figure}
\epsscale{0.75}
\plotone{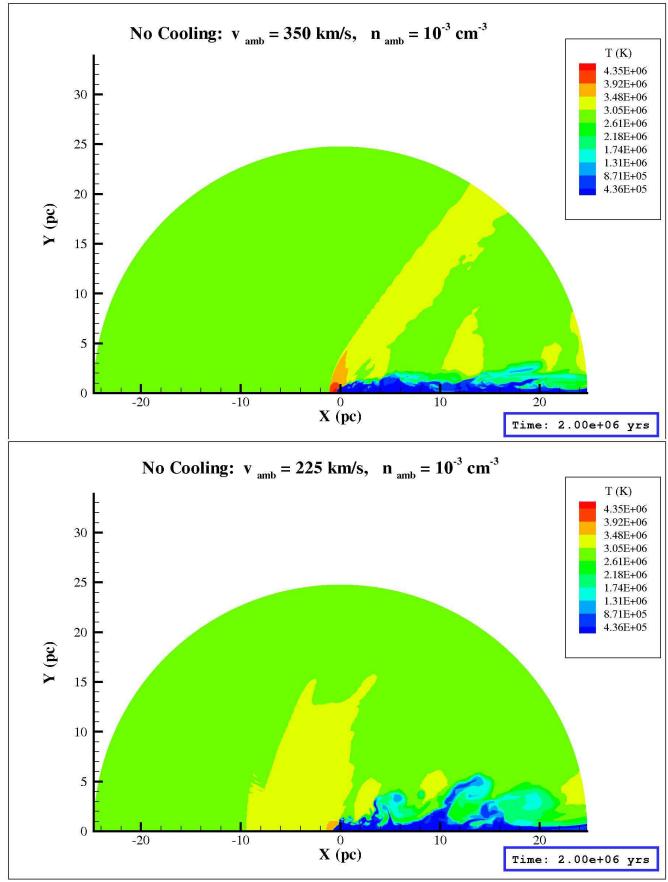}
\caption[Temperature maps showing the influence of a low stellar velocity]%
        {Temperature maps showing the influence of a low stellar velocity, at a
        simulation time of $2 \times 10^6$ years (runs FDFV and FDLV).   
        The flow in the FDLV case
        (bottom panel) has similar large mode instability properties as the 
	fiducial case (top panel; FDFV), but there are fewer small
	modes to aid in the wake mixing.  The prominent bow shock
	is absent.}
\label{fig:lvfdcompfull}
\end{figure}

In this simulation (FDLV), the lower stellar velocity of $v_{amb} = 225$ km s$^{-1}$ results 
in a subsonic ambient flow with Mach number $M \approx 0.87$ (we will term this 
simulation the  subsonic case, although the ambient gas is still supersonic 
with respect to the stellar ejecta). 
The subsonic stellar velocity leads to the absence of a bow shock, which was present
in all other simulations (Figure~\ref{fig:lvfdcompfull}).
If the mass-losing star were replaced with a solid body, laminar flow around the object
would be described by potential flow without any shocks.  However, the interaction
between the mass loss from the star and the flow around it is not steady in time as
the two regions interact to produce a wake and create KH instabilities.
These differences relative to the flow around a solid body lead to the development of 
weak shocks in a flow where the Mach number is close to unity.

\begin{figure}
\epsscale{0.75}
\plotone{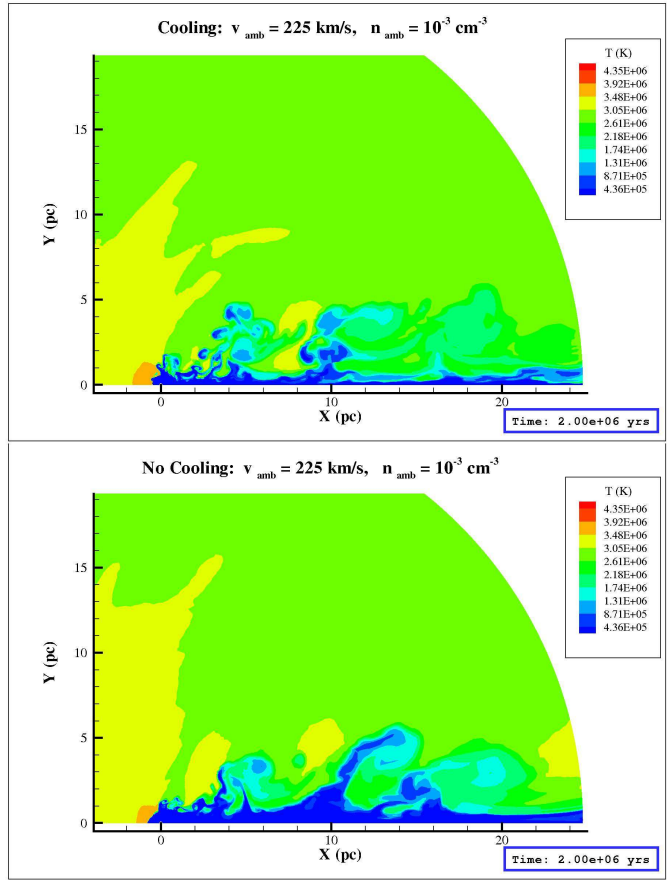}
\caption[Temperature maps showing the influence of cooling on FDLV]%
        {Temperature maps showing the influence of cooling on the low velocity
        simulations (FDLVC and FDLV), at a
        simulation time of $2 \times 10^6$ years.   The cooled flow (top panel)
        is more unstable than the adiabatic wake (bottom panel).  The prompt
	mixing by instabilities near the head of the cooled flow is once again
	seen in this case.  The cooled wake is more narrow, as expected.}
\label{fig:lvcoolcomp}
\end{figure}

\begin{figure}
\epsscale{1.0}
\plotone{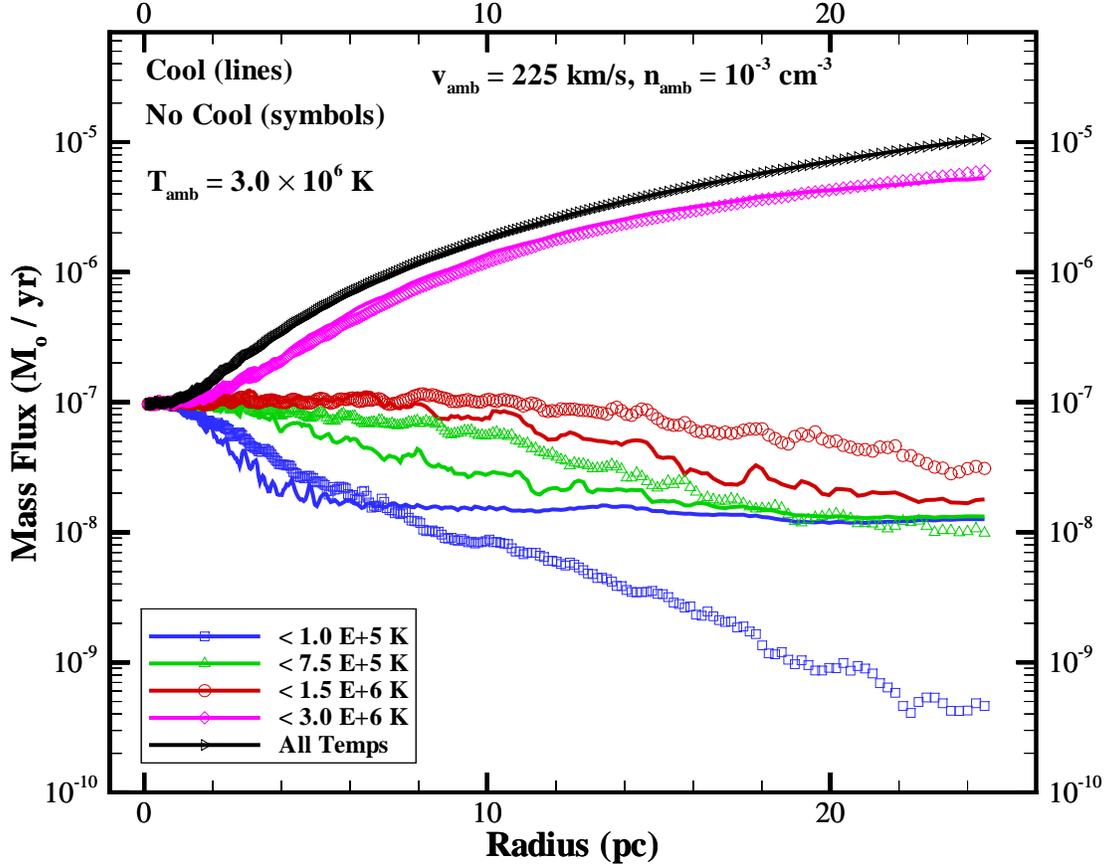}
\caption[FDLV mass flux with and without cooling]%
        {The mass fluxes in the low velocity simulations with and without 
        radiative cooling (FDLVC and FDLV).  This is very
	similar to the fiducial comparison.  The cooled flow appears more
	unstable, but any stellar ejecta that escapes the heating near the head 
	of the flow is locked into cold blobs and the narrow
	wake.  The adiabatic wake is continuously mixed and heated in the
	wake as it flows downstream.  The small gap between the ambient and all 
	temperature profiles
	indicates that a relatively small amount of bow shock heating occurs.}
\label{fig:mffdlvc-nc}
\end{figure}

In simulations with or without radiative cooling (FDLVC and FDLV), 
weak shocks develop nearly 10 pc upstream.  
Kelvin-Helmholtz instabilities grow rapidly, being especially prominent in the 
several parsec downstream region (Figure~\ref{fig:lvcoolcomp}).
In the non-cooling case, this leads to efficient heating so that when the wake
exits the simulation, the gas cooler than 10$^5$ K is about 0.005 of the initial
mass flux (Figure~\ref{fig:mffdlvc-nc}).
When radiative cooling is introduced (FDLVC), there is more active development of KH
instabilities that peel off near the head of the flow and in the 10 pc region
downstream in the wake (Figure~\ref{fig:lvcoolcomp}).  Another difference,
also common with other radiative simulations, is that the cool wake is 
thinner as cool material is more easily compressed.  This modifies the axial
ratio of the wake, and evidently, the dominant KH growth modes.

An expectation of this lower velocity flow was that the heating would be less
effective than in the fiducial run, so more gas would remain cool when radiatives
losses were included (run FDLVC).  However, calculations fail to support this expectation and
the amount of cooled gas is not greater than in the fiducial run and may be slightly
less.  In the last 10 pc before the gas leaves the grid, the mass flux of cooled gas
has reached asyptotic values at 0.14 of the initial mass flux for gas cooler than
either 10$^5$ K or $7.5 \times 10^5$ K.  Even including gas cooler than 
$1.5 \times 10^6$ K does not raise the fraction above 0.2 of the initial mass flux.
This decrease in the cool mass flux, relative to the fiducial run, is most likely
due to the more effective mixing and small-scale shocks that occur in this subsonic
simulation.

\section{Discussion and Conclusions}

We have, for the first time, calculated in detail the interaction between 
stellar mass loss and the hot interstellar medium in elliptical galaxies using
accurate two-dimensional numerical hydrodynamics simulations.  The
ultimate fate of the stellar ejecta has significant implications for the 
standard theory of the interstellar medium in elliptical galaxies, because this
mass loss is the primary origin of the $\sim 1$ keV X-ray gas in these galaxies.
We examined whether the optically thin radiative cooling of the gas
will prevent some stellar ejecta from being shock heated to the local
velocity dispersion temperature and entering the ambient flow.
Simple calculations suggest that the fraction of stellar mass loss 
that remains cool might depend strongly 
the density of the surrounding and the velocity of the star relative to 
the hot ambient interstellar medium (Table 2).
For the simulations of this process, we calculated a reference (fiducial) case against
which changes in density and stellar velocity were compared.  
Different ambient densities are naturally encountered in early-type galaxies,
where the density is considerably larger in the inner part compared to the
outer region.  Variations in the stellar velocity occur in the orbit of every star.
For each set of initial conditions, two simulations were calculated, 
one with and one without radiative losses.
The gas exiting the grid may have a mean temperature larger or smaller than the 
initial ambient temperature.  It will have either greater or less entropy than
the surroundings, and in pressure equilibrium with its surroundings, will be 
positively bouyant and rise outward in the galaxy, or flow inward.
We will not consider this complication here, as it occurs on a timescale longer
than the flow time of the calculations.

For the runs where the stellar velocity was supersonic, it had a Mach number
of $M = 1.3 - 1.8$, and this led to a bow shock in the ambient medium at about 
the theoretically expected standoff distance, $\sim 1$ pc upstream.
Whereas the stellar mass loss is also shocked, this is not the event that
is the primary heating mechanism for the cool gas, although most of the heating 
occurs in the region near the star, within $\sim 3$ pc downstream.
The initial interaction of the streaming ambient ISM with the stellar ejecta 
imparts momentum to the ejecta that begins to drive it downstream.
The flow past the cold ejecta is Kelvin-Helmholtz unstable, and fingers
of material are drawn out perpendicular to the intial flow axis.  
This greatly increases the cross section of the stellar ejecta, leading
to heating by shocks that are driven into the cooler material by the rapidly flowing ambient
medium.  The other important mechanism is the mixing between the fast-flowing
hot gas and the stellar ejecta, which occurs through the presence of vortices
and the development of Rayleigh-Taylor instabilities.
These instabilities and the related heating occur further down the wake as well,
and in the absence of radiative cooling, nearly all of the gas has been heated
and will eventually become part of hot ambient medium.  We constructed our
fixed grid so that nearly all of the important processes occur before the 
flow exits the grid (25 pc downstream from the star).  
The timescales proved to be longer for the low-density run
(LDFV, LDFVC), where the cool material is still decreasing as the gas exits the grid.

When radiative cooling is included, and when the cooling time of the gas is 
small enough to be dynamically important, there is an increase in the
upstream mixing and heating of the ejecta by instabilities, but any material 
that is not promptly heated near the head of the interaction condensed into cold 
blobs or flowed onto the central wake.
There continues to be instabilities between the cold wake and the hotter and
faster flow, which produces a net heating rate.  There is also a radiative
cooling rate associated with the wake and if the evolution of the wake
depends strongly upon whether this is larger or smaller than the heating process.
Also, the heating due to shocks and mixing decreases with downstream distance,
so the two can eventually balance or the cool component in the wake can even
grow if it cools the hotter material as it mixes in.
This appears to be the case in the fiducial run, where $15\%$ of the gas 
is cold ($ < 10^5$ K) when it exits the grid.  However, as the flow approaches
the last 10 pc of the grid, the mass flux fraction below $10^5$ K is rising
as the fraction of gas in the $10^5$ - $1.5 \times 10^6$ K range is falling.
Most of the gas in the $10^5$ - $1.5 \times 10^6$ K range appears to be heated
and join the hot ambient medium, but some fraction grows onto the wake.
We estimate that about $20\%$ of the gas might remain cold if the simulation could
be extended to larger radii.

We expected the density of the ambient medium to be crucial in determining
the amount of stellar ejecta that remained cool.  This is because the
wake material tries to approach pressure equilibrium with the hot ambient
flow.  Consequently, a higher ambient density gives rise to an increased density
in the cooler wake, and at a given temperature, the radiative cooling rate 
increases as the square of the density.  However, the higher density simulation shows that
while a high density (pressure) entrains the wake more, it also
magnifies the strength of the Kelvin-Helmholtz instabilities as they mix
and shock the inner wake material.  This leads to an increased heating
rate that partly offsets the increased cooling rate.
The simulations show that there is an increase in the amount of stellar ejecta
that remains cool, but that increase is rather modest.  Where the flow exits
the grid, $25\%$ of the initial stellar mass loss is cold ($ < 10^5$K), $5-10\%$
more than in the fiducial run.  Even in the low-density run, there appears
to be some cold material remaining, although those mass flux values have not reached
stable asymptotic limits and appear to be decreasing.  We regard the nominal
mass flux for the cold gas of $19\%$ as an upper limit and not a reliable value.

Variation of the stellar velocity naturally changes the temperature of the 
shocked material, where higher stellar velocities produce gas at a higher temperature.
For the higher velocity case, we expected greater heating and shorter growth times
for the KH instability, as the instability is inversely proportional to the relative flow velocity.
These combined effects led to significantly less cool gas remaining in the wake,
where $5.7\%$ of the initial stellar mass loss remained cold ($ < 10^5$K) as it exited
the grid, but this fractional amount is still slowly declining.  
Compared to the fiducial case, faster moving stars have their mass loss more 
efficiently heated to the ambient gas temperature.

The opposite expectation, that more gas will remain cool for lower velocity stars,
was not supported by simulations.  The lower stellar velocity, 225 km s$^{-1}$, has a Mach
number slightly below unity, so a bow shock is not produced.  However, the interaction
with the stellar mass loss leads to time-dependent structures that produce shocks in
the stellar mass loss as well as in the ambient hot gas.  For the runs without radiative
cooling, there is more rapid heating of the stellar mass loss through shocks and mixing,
compared to the fiducial case.  For the runs with radiative cooling, most of the heating
occurs even more rapidly within 2 pc of the star, but for distances further downstream,
radiative cooling exceeds heating for the remaining cool gas ($ < 10^5$K).
The final amount of cool gas exiting the grid is similar to, but slightly less than 
the amount in the fiducial run.

One of the surprises of the simulations is that there are dominant modes in the 
growth of Kelvin-Helmholtz instabilities.  The classical calculation has no
preferred wavelength, so the smallest modes grow first, as they have the fastest growth rate.
However, that calculation assumes the the two fluids have an infinite size and 
issues such as radiative cooling are not present.
In these calculations, the wake that is produced has a characteristic size and 
for the radiative flows, there is a characteristic cooling time as well.
\citet{hard97} studied the related situation of radiatively cooling jets and
found the growth of preferred modes arising from Kelvin-Helmholtz instabilities \citep{xu2000}.
The preferred growth modes found in our simulations may arise from the same
physical basis and we will analyze this aspect in an upcoming work.

There are a number of shortcomings in these simulations and various issues that 
still need to be addressed.  One improvement 
would be to employ a full three-dimensional geometry.  This
will eliminate the numerical problems that accompany an axisymmetric
geometry.  Also, the number of possible instability modes is likely to
increase in three dimensions, which would affect the mixing of the wake 
and the amount of gas that would ultimately remain cool (\citealt{brmfmr98}),
although experience has show that it can be difficult predicting the 
outcome of moving to three-dimensional calculations with any accuracy.
Another improvement would be to extend the grid and better resolve regions of
enhanced activity through subgridding.

A third possible improvement is the inclusion of magnetic fields.  
The inner regions of elliptical galaxies have small 
magnetic field strengths of $B \sim 1 \mu{\rm G}$ \citep{mb97}, and the 
standard argument against
considering magnetic fields in simulations such as ours is that the kinetic
energy density of the gas is several orders of magnitudes larger than the
magnetic energy density:  $U_{KE}/U_{B} \sim 10^{3}$.  Most studies of magnetic
fields in elliptical galaxies concern possible dynamo mechanisms to create
the stronger $\sim 50 \mu{\rm G}$ fields seen in galactic halos \citep{lb90,ms96,mb97}.  
However, work by \citet{jones97} on the
role of very weak magnetic fields in the evolution of vortices generated
by Kelvin-Helmholtz instabilities indicates that the dissipative influence
of magnetic fields on such instabilities should not be ignored.  
It is clear that the twisting and compression of fields in the stellar wind
can make magnetic fields important.  The general effect
of magnetic fields would be to stabilize the wake against instabilities, 
thereby reducing the degree of heating and mixing.
The strength and topology of magnetic fields will help determine the importance of 
conduction and viscosity.  Given the complicated mixing flows, the downstream
magnetic field is likely to be tangled, which supresses conduction.

Another parameter that should be investigated is the metallicity of the 
mass lost from the star.  The metallicity of the gas is important for the
radiative cooling in this temperature range, as most of the cooling is due to
atomic lines.  Higher metallicity leads to shorter cooling times and presumably
a greater fraction of the stellar mass lost remains cool.  This would lead to
lower metallicity gas preferentially being added to the hot medium.

Provided that more sophisticated calculations do not change the basic result of
this work, our simulations imply that there would be small amounts of cool gas 
distributed around a galaxy.  There would be a preference for such material to
be more common in regions of higher ambient gas density, which occurs in the inner
parts of the galaxy.  For a given ambient density, the cooled gas would preferentially
originate from stars with elevated metallicities and that are not on the high 
velocity part of the Maxwellian distribution.
One implication is that there would be cool gas available for photoionization
by the UV-rising branch of the spectral energy distribution of early-type galaxies.
Warm ionized gas is frequently detected from early-type galaxies, and in 
many cases, the spatial distribution is similar to that of the stars, consistent
with the predictions from our simulations \citep{caon00}.
Although the distribution of the optical emission line gas is consistent with our
expectations, the velocity structure is often different than expected.
We would predict that the cooled gas should have the same rotational velocity 
axis, yet some fraction of the galaxies show the gas to be counter-rotating
when compared to the stars.  At least in those galaxies, an external origin for 
the gas seems more likely \citep{caon00}.

Our calculations follow the wakes during their initial heating and cooling stages, 
but the cooled wakes lack the buoyancy of the hotter gas, so they will eventually
sink inward.  The interaction of the wakes with their surroundings during this
inflow stage is beyond the scope of our calculations.  The material may ultimately
be heated and mixed into the ambient medium, or act as nucleation sites
for further cooling.
Gas and dust are seen in the inner parts of early-type galaxies and this could be
the end result of cooled wake material that has flowed inward \citep{temi07}.

A complementary process is that there is a fraction of the hot ambient gas that is
heated above its initial value.  This hotter gas will be buoyant and rise outward in the
galaxy.  This heating of the ambient gas may be one of the sources that leads
to the ambient gas being above the velocity dispersion temperature of the galaxy.

There is a bias in that the mass loss from higher velocity stars is more readily 
heated.  As a consequence, the temperature of the hot ambient medium would be
greater than the velocity dispersion temperature, even in the absence of any
supernova heating.  Extrapolating from our results, this would cause the ambient
temperature to lie $\sim 10\%$ above the velocity dispersion temperature.

Finally, there are two modes of mass loss that were not considered here:  
planetary nebulae and supernovae.  A planetary nebulae is an event that is more 
discrete in time, so it presents different challenges to numerical modeling.  
We have worked on this issue and it will be presented in a future work.
We have no plans to study numerically the evolution of supernovae in the hot
environment found in an early-type galaxy.

\acknowledgements
JRP would like to give special thanks to Philip Hughes and Hal Marshall for their
valuable advice and comments during code development, simulations, and on the
content of my PhD Thesis.  Also, thanks are due to Gus Evrard and Douglas Richstone
for serving on JRP's thesis committee and for providing timely support and advice.
Financial support is gratefully acknowledged and was provided to JRP through
a Department of Energy Computational Science Graduate Fellowship and to JNB through
a Long Term Space Astrophysics Grant from NASA.

\clearpage

\begin{deluxetable}{rcccccc}
\tablecolumns{7}
\tablewidth{0pc}
\tablecaption{Simulation Input Parameters}
\tablehead{
   \colhead{} & \colhead{} & \colhead{Ambient Medium} & \colhead{} & \colhead{} & \colhead{Stellar Ejecta} & \colhead{}\\
   \colhead{Model} & \colhead{n} & \colhead{v} & \colhead{T} 
   & \colhead{n} & \colhead{v} & \colhead{T}                   \\
   \colhead{} & \colhead{($10^{-3}$ cm$^{-3}$)} & \colhead{(km s$^{-1}$)} & \colhead{($10^6$ K)} 
      & \colhead{($10^{-3}$ cm$^{-3}$)} & \colhead{(km s$^{-1}$)} & \colhead{($10^6$ K)} }
\startdata
FDFV & 1 & 350 & 3 & 1 & 35 & 1 \\
HDFV & 3.33 & 350 & 3 & 1 & 35 & 1 \\
LDFV & 0.1 & 350 & 3 & 1 & 35 & 1 \\
FDHV & 1 & 475 & 3 & 1 & 35 & 1 \\
FDLV & 1 & 225 & 3 & 1 & 35 & 1 \\ 
\enddata
\tablecomments{(F)iducial, (H)igh, (L)ow, (D)ensity, (V)elocity} 
\end{deluxetable}

\begin{deluxetable}{ccccc}
\tablecolumns{5}
\tablewidth{0pc}
\tablecaption{Fraction of Cold and Ward Mass Ejecta Remaining}
\tablehead{
   \colhead{} & \colhead{Non-radiative} & \colhead{Non-radiative} & \colhead{Radiative} & \colhead{Radiative} \\
   \colhead{Model} & \colhead{$T \leq 10^5$ K} & \colhead{$T \leq 7.5 \times 10^5$ K}
    & \colhead{$T \leq 10^5$ K} & \colhead{$T \leq 7.5 \times 10^5$ K} \\ }
\startdata
FDFV & $1.4 \times 10^{-3}$ & $0.40$ & $0.15$ & $0.19$ \\
HDFV & $0$ & $0.18$ & $0.25$ & $0.26$ \\
LDFV & $0.11$ & $0.59$ & $0.19$ & $0.58$ \\
FDHV & $1.1 \times 10^{-2}$ & $0.62$ & $5.7 \times 10^{-2}$ & $0.16$ \\
FDLV & $4.6 \times 10^{-3}$ & $0.10$ & $0.14$ & $0.14$ \\
\enddata
\end{deluxetable}

\clearpage

\clearpage

\clearpage

\clearpage

\clearpage

\end{document}